\documentclass[11pt]{article}

\usepackage[letterpaper, text={6.5in, 9.0in}]{geometry}
\usepackage{amsmath}
\usepackage{amssymb}
\usepackage{amsthm}
\usepackage{xcolor}
\usepackage{graphicx}

\newtheorem{proposition}{Proposition}

\colorlet{revision}{black}

\begin{document}

\title{A Bayesian Approach for Robust Longitudinal Envelope Models}
\author{Peng Zeng and Yushan Mu \\ 
Department of Mathematics and Statistics, Auburn University}


\maketitle

\begin{abstract}
The envelope model provides a dimension-reduction framework for multivariate linear regression. However, existing envelope methods typically assume normally distributed random errors and do not accommodate repeated measures in longitudinal studies. To address these limitations, we propose the robust longitudinal envelope model (RoLEM). RoLEM employs a scale mixture of matrix-variate normal distributions to model random errors, allowing it to handle potential outliers, and incorporates flexible correlation structures for repeated measurements. In addition, we introduce new prior and proposal distributions on the Grassmann manifold to facilitate Bayesian inference for RoLEM. Simulation studies and real data analysis demonstrate the superior performance of the proposed method.
\end{abstract}

\section{Introduction}

In longitudinal studies, multivariate responses are often collected repeatedly from the same subjects over time. Beyond modeling the means and variances of these responses, it is crucial to account for the correlations of the same response across time as well as the correlations among different responses at a given time point. As the number of responses increases, the number of parameters grows quadratically. Efficient modeling strategies are therefore essential for capturing and interpreting these complex dependence structures.

The envelope model is a dimension-reduction framework for multivariate linear regression \cite{cook2010envelope}. 
It assumes that the response vector can be decomposed into two components: 
one that depends on the covariates and one that does not. 
Estimation efficiency can be improved by identifying the component associated with the covariates. 
The envelope method has been extended to numerous settings 
\cite{cook2015foundations, ding2018matrix, ding2021envelope, cook2018introduction, lee2020review}. 
A brief review is provided in Section~\ref{subsection:envelope}.

Existing envelope models, which typically assume independence among observations, are not directly applicable to longitudinal studies. By ignoring within-subject correlation, these models may yield inaccurate estimates of variability and lead to flawed statistical inference. More recently, a mixed-effects envelope model has been proposed to incorporate random effects, and an EM algorithm has been developed for model fitting \cite{shi2020mixed}.

Bayesian methods for envelope models with independent observations have been studied previously \cite{khare2017bayesian, chakraborty2023comprehensive}, which represent two different parameterizations of the covariance matrix; 
see Section~\ref{subsection:envelope} for details. 
To this end, we aim to develop a Bayesian framework for envelope models to analyze longitudinal data. 
Additionally, we introduce a new prior distribution that facilitates the incorporation of existing knowledge, 
together with a novel proposal distribution that enables efficient sampling within a Metropolis-Hastings algorithm.

\textcolor{revision}{The assumption of normal random errors} is commonly adopted in existing envelope models for theoretical and computational convenience. However, this assumption makes the models sensitive to outliers and heavy-tailed error distributions. There is therefore a need to relax \textcolor{revision}{the assumption of normal random errors} to enhance the robustness and flexibility of envelope models.

In this paper, we propose a novel Bayesian approach for robust longitudinal envelope models (RoLEM). 
This approach has three key features: 
it accommodates various correlation structures for repeated measures, 
it relaxes the \textcolor{revision}{assumption of normal random errors} by employing a scale mixture of matrix-variate normal distributions to achieve robustness, 
and it introduces a flexible prior that readily incorporates existing knowledge. 
Simulation studies and real data analysis demonstrate the superior performance of the proposed RoLEM. 
\textcolor{revision}{The most comparable existing method is the mixed-effects envelope model \cite{shi2020mixed}. However, it is equivalent to RoLEM only under a compound-symmetric correlation structure, and it neither extends to other correlation structures nor accounts for potential outliers.}

The remainder of the paper is organized as follows.
Section~\ref{section:review} reviews envelope models and related distributions.
Section~\ref{section:model} presents the model specification in detail and derives the posterior sampling algorithm.
Section~\ref{section:implementation} discusses implementation issues, including 
a Metropolis-Hastings algorithm, choices of hyperparameters, initialization,  model selection, and software.
Section~\ref{section:example} illustrates the proposed methods through simulation studies and a real data analysis.
Section~\ref{section:conclusion} concludes the paper with additional discussion.
Additional proofs, tables, and figures are provided in the supplementary material.

\section{Review of Envelope Models and Related Distributions}
\label{section:review}

This section provides a brief review of relevant literature. 
Section~\ref{subsection:envelope} reviews envelope models;
Section~\ref{subsection:scale-mixture} reviews scale mixtures of matrix-variate normal distributions; and
Section~\ref{subsection:Grassmann} reviews distributions on the Grassmann manifold.

\subsection{Envelope models}
\label{subsection:envelope}

Consider the multivariate linear regression model
\begin{align*}
y = \alpha + \beta x + \varepsilon,
\end{align*}
where $y \in \mathbb{R}^{r}$ is the response vector, $x \in \mathbb{R}^{p}$ is the covariate, $\alpha \in \mathbb{R}^{r}$ and $\beta \in \mathbb{R}^{r \times p}$ are unknown parameters, and $\varepsilon \in \mathbb{R}^{r}$ is a random error.
For an $r \times r$ orthogonal matrix $(\Gamma, \Gamma_{0})$, 
where $\Gamma \in \mathbb{R}^{r \times u}$ and $\Gamma_{0} \in \mathbb{R}^{r \times (r-u)}$, 
the envelope model \cite{cook2010envelope} assumes that
(a) $\Gamma_{0}^{T}y \mid x \sim \Gamma_{0}^{T}y$, where $\sim$ denotes identical distribution, and
(b) $\Gamma^{T}y$ is uncorrelated with $\Gamma_{0}^{T}y$ given $x$.
This implies that the response vector $y$ can be decomposed into two components: 
one that depends on the covariates $x$ and one that does not. 
Under this assumption, the regression coefficient $\beta$ and the error covariance matrix 
$\Sigma_{\varepsilon} = \text{var}(\varepsilon) \in \mathbb{R}^{r \times r}$ are structured as
\begin{align*}
\beta = \Gamma \eta, \quad
\Sigma_{\varepsilon} = \Gamma \Omega \Gamma^{T} + \Gamma_{0} \Omega_{0} \Gamma_{0}^{T},
\end{align*}
where $\eta \in \mathbb{R}^{u \times p}$, 
$\Omega \in \mathbb{R}^{u \times u}$, 
and $\Omega_{0} \in \mathbb{R}^{(r-u) \times (r-u)}$. 
The integer $u$ is referred to as the dimension of the envelope model.

The parameters to be estimated are 
$\theta = \{\alpha, \beta, \Sigma_{\varepsilon}\} = \{\alpha, \eta, \Gamma, \Gamma_0, \Omega, \Omega_0\}$. 
Both Khare et al. \cite{khare2017bayesian} and Chakraborty and Su \cite{chakraborty2023comprehensive} 
recommend a non-informative uniform prior for $\alpha$ and a normal prior for $\eta \mid \Gamma$. 
Constraints on ${\Gamma, \Gamma_0, \Omega, \Omega_0}$ are required to ensure identifiability. 
The two studies adopt different strategies for imposing these constraints, resulting in distinct prior specifications.

Khare et al. \cite{khare2017bayesian} assume that $(\Gamma, \Gamma_0)$ forms an orthogonal matrix and that $\Omega$ and $\Omega_0$ are diagonal matrices with ordered positive entries. Under this parameterization, the $r(r+1)/2$ parameters in $\Sigma_{\varepsilon}$ are represented by $r(r-1)/2$ parameters in $(\Gamma, \Gamma_0)$ and $r$ parameters in $\{\Omega, \Omega_0\}$. The prior distribution for $(\Gamma, \Gamma_0)$ is a matrix-variate Bingham distribution, while the diagonal entries of $\Omega$ and $\Omega_0$ are assigned independent inverse-Gamma priors. 
Khare et al. \cite{khare2017bayesian} also develop an algorithm for posterior sampling, although it is complex and challenging to implement.

Chakraborty and Su \cite{chakraborty2023comprehensive} parameterize $(\Gamma, \Gamma_0)$ using a matrix $A \in \mathbb{R}^{(r-u) \times u}$. Specifically, for any given $A$, one can construct column-orthogonal matrices
$\Gamma = C_A(C_A^TC_A)^{-1/2}$ and $\Gamma_0 = D_A(D_A^TD_A)^{-1/2}$ 
satisfying $\Gamma^T \Gamma_0 = 0$, where $C_A = (I_u, A^T)^T$ and $D_A = (-A, I_{r-u})^T$.
They further assume that $\Omega$ and $\Omega_0$ are positive definite. Under this parameterization, the $r(r+1)/2$ parameters in $\Sigma_{\varepsilon}$ are represented by $u(r-u)$ parameters in $(\Gamma, \Gamma_0)$ or equivalently $A$, $u(u+1)/2$ parameters in $\Omega$, and $(r-u)(r-u+1)/2$ parameters in $\Omega_0$.
Chakraborty and Su \cite{chakraborty2023comprehensive} assign a matrix-variate normal prior to $A$ and independent inverse-Wishart priors to $\Omega$ and $\Omega_0$. While computationally convenient, using a normal prior for $A$ makes it difficult to incorporate prior knowledge about $(\Gamma, \Gamma_0)$.

The parameterization of our method is similar to that of Chakraborty and Su \cite{chakraborty2023comprehensive}. However, instead of placing a prior on $A$, we directly specify a prior on $\Gamma$, or more precisely, on $P = \Gamma \Gamma^T$. Note that there is a one-to-one correspondence between $A$ and $P = \Gamma \Gamma^T$. Since $P$ is idempotent and uniquely defines a linear subspace, this establishes a one-to-one correspondence between $A$ and $u$-dimensional linear subspaces, which form a Grassmann manifold. By specifying the prior directly on the Grassmann manifold, it becomes straightforward to incorporate existing knowledge about $(\Gamma, \Gamma_0)$. Section~\ref{subsection:Grassmann} provides a brief review of the Grassmann manifold.

Both Khare et al. \cite{khare2017bayesian} and Chakraborty and Su \cite{chakraborty2023comprehensive} assume that the random errors are independently distributed according to a multivariate normal distribution. In this paper, we instead assume that the random errors follow a scale mixture of matrix-variate normal distributions, a flexible class capable of capturing outliers. Moreover, our approach accommodates correlation structures arising from repeated measurements.

\subsection{Scale mixture of \textcolor{revision}{matrix-variate} normal distributions}
\label{subsection:scale-mixture}

This subsection introduces the scale mixture of matrix-variate normal (SMMN) distributions, which extend the normal distribution to accommodate outliers. In addition, a stochastic representation of the SMMN distribution provides a hierarchical model that can be used in practice for sampling.

Let $M \in \mathbb{R}^{a \times b}$ be an arbitrary matrix, 
and let $\Lambda_1 \in \mathbb{R}^{a \times a}$ and $\Lambda_2 \in \mathbb{R}^{b \times b}$ be positive definite matrices. 
A random matrix $Y \in \mathbb{R}^{a \times b}$ is said to follow a matrix-variate normal distribution \cite{GuptaNagar2000}, denoted by $\text{MN}(M, \Lambda_1, \Lambda_2)$, if its density is
\begin{align*}
    (2\pi)^{-ab/2}
	|\Lambda_{1}|^{-b/2} |\Lambda_{2}|^{-a/2}
	\text{etr}\left(
	-\frac{1}{2}
	\Lambda_{1}^{-1}(Y - M)\Lambda_{2}^{-1}(Y - M)^{T}
	\right),
\end{align*}
where $\text{etr}(X) = \exp(\text{tr}(X))$ for a matrix $X$.
\textcolor{revision}{$\Lambda_1$ and $\Lambda_2$ are referred to as the row and column covariance matrices, respectively, 
and they capture the covariance structures of the rows and columns of $Y$.}
It can be shown that $Y \sim \text{MN}(M, \Lambda_1, \Lambda_2)$ if and only if
$\text{vec}(Y)\sim N(\text{vec}(M), \Lambda_{2}\otimes \Lambda_{1})$,
where $\text{vec}(\cdot)$ stacks the columns of a matrix into a vector, and $\otimes$ denotes the Kronecker product.

Let $G(\cdot; \nu)$ be a univariate probability distribution with parameter $\nu$ and support on $(0, \infty)$. 
A random matrix $Y \in \mathbb{R}^{a \times b}$ is said to follow a scale mixture of matrix-variate normal distributions, denoted by $\text{SMMN}(M, \Lambda_1, \Lambda_2, G)$, if its density is
\begin{align*}
	(2\pi)^{-ab/2}|\Lambda_1|^{-b/2}|\Lambda_2|^{-a/2}
    \int_{0}^{\infty}
	u^{ab/2}\text{etr}\left(-\frac{u}{2}\Lambda_1^{-1}(Y - M)\Lambda_2^{-1}(Y - M)^T\right) dG(u; \nu).
\end{align*}
A stochastic representation of the SMMN distribution \cite{azzalini2014skew} is
\begin{align*}
	Y = M + \tau^{-1/2}Z,
\end{align*}
where $Z \in \mathbb{R}^{a \times b} \sim \text{MN}(0, \Lambda_1, \Lambda_2)$ is independent of $\tau \sim G(\cdot; \nu)$. This leads to the hierarchical representation
\begin{align*}
	Y\mid \tau \sim \text{MN}(M, \tau^{-1}\Lambda_1, \Lambda_2), \quad \tau \sim G(\cdot; \nu),
\end{align*}
for which the marginal distribution of $Y$ is exactly $\text{SMMN}(M, \Lambda_1, \Lambda_2, G)$.
The first two central moments of $Y$ are
\begin{align*}
  E(Y) &= E[E(Y\mid\tau)] = M, \\
  var(\text{vec}(Y))
  &= var[E(\text{vec}(Y)\mid \tau)] + E[var(\text{vec}(Y)\mid \tau)]
   = (\Lambda_2\otimes\Lambda_1) \cdot E(\tau^{-1}).
\end{align*}

By choosing an appropriate distribution $G(\cdot; \nu)$, the SMMN distribution can accommodate outliers. 
For example, when $\tau$ is a finite mixture of point masses, $Y$ becomes a finite mixture of normal distributions. 
When $\tau \sim \text{Gamma}(\nu/2, \nu/2)$, $Y$ follows a matrix-variate $t$-distribution \cite{GuptaNagar2000}, denoted by $\text{MT}(\nu, M, \Lambda_1, \Lambda_2)$, with density
\begin{align*}
    \frac{\Gamma((ab + \nu)/2)}{\Gamma(\nu/2) \nu^{ab/2} \pi^{ab/2}}|\Lambda_1|^{-b/2}|\Lambda_2|^{-a/2}
    \left[ 1 + \frac{1}{\nu}
    \text{tr}\left(\Lambda_1^{-1}(Y - M)\Lambda_2^{-1}(Y - M)^T\right)
    \right]^{-(ab + \nu)/2},
\end{align*}
where $\nu$ is the degrees of freedom and $\Gamma(\cdot)$ denotes the Gamma function. 
When $\nu > 1$, the mean of $Y$ is $M$, and 
when $\nu > 2$, the variance of $\text{vec}(Y)$ is $(\nu/(\nu - 2)) (\Lambda_2 \otimes \Lambda_1)$.

In this paper, we present the general model using the SMMN distribution, but focus on the matrix-variate $t$-distribution in our algorithms and examples due to its wide applicability. 
Extensive studies on $t$-regression have been conducted in literature 
\cite{fernandez1999multivariate, fonseca2008objective, wang2012bayesian, ding2014bayesian}.

\subsection{Distributions on Grassmann manifold}
\label{subsection:Grassmann}

This subsection introduces several distributions on the Grassmann manifold. For a comprehensive discussion, see Chikuse \cite{chikuse2003statistics}. Here, we focus only on the distributions relevant to this paper.

The Grassmann manifold $\mathcal{G}_{m, k}$ is the space of all $k$-dimensional linear subspaces of $\mathbb{R}^m$. 
Each $k$-dimensional subspace corresponds uniquely to an $m \times m$ idempotent matrix of rank $k$, 
so the Grassmann manifold $\mathcal{G}_{m, k}$ can be equivalently represented by 
\begin{align*}
	\mathcal{P}_{m, k}
	= \{P\in\mathbb{R}^{m\times m} \mid PP = P, \; \text{rank}(P) = k\}.
\end{align*}
We typically write $P = \Gamma \Gamma^T$, where the columns of $\Gamma$ form an orthogonal basis for the subspace corresponding to $P$. The choice of $\Gamma$ is not unique: for any $k \times k$ orthogonal matrix $U$, $\Gamma U$ spans the same subspace as $\Gamma$, because $P = (\Gamma U)(\Gamma U)^{T} = \Gamma\Gamma^{T}$.



We introduce two distributions on $\mathcal{P}_{m, k}$ that exhibit a concentration property, analogous to how a normal distribution is concentrated around its mode.
Note that the density function of a distribution on $\mathcal{P}_{m, k}$ is defined with respect to the Haar measure or uniform distribution on $\mathcal{P}_{m, k}$.
The matrix Langevin distribution on $\mathcal{P}_{m, k}$ \cite{chikuse2003statistics} has density
\begin{align}
	f(P)
	= \frac{1}{_{1}F_{1}(\frac{1}{2}k; \frac{1}{2}m; M)}
	  \text{etr}(MP),
\label{eq:Langevin}
\end{align}
where $_{1}F_{1}(\cdot; \cdot; \cdot)$ is a hypergeometric function with a matrix argument, and $M \in \mathbb{R}^{m \times m}$ is symmetric. 
This distribution is closely related to the matrix Bingham distribution: if $\Gamma \in \mathbb{R}^{m \times k}$ is a random column-orthogonal matrix following a matrix Bingham distribution with parameter $M$, then the density of $P = \Gamma \Gamma^T$ is exactly (\ref{eq:Langevin}).
Another common distribution on $\mathcal{P}_{m, k}$ \cite{chikuse2003statistics} has density
\begin{align}
  	f(P) &= |M|^{-k/2} |I_m - P + M^{-1}P|^{-m/2},
\label{eq:angular}
\end{align}
where $M \in \mathbb{R}^{m \times m}$ is symmetric. 
This distribution characterizes the linear space spanned by a normal random matrix.
The connection between the distribution (\ref{eq:angular}) and the normal distribution makes it particularly suitable as a proposal distribution in a Metropolis-Hastings algorithm, since sampling from it is straightforward.


\color{revision}
\begin{proposition}
If $Z$ is an $m \times k$ random matrix following a matrix-variate normal distribution $\text{MN}(0, M, I_k)$, then the corresponding projection matrix $P = Z (Z^T Z)^{-1} Z^T$ has the density given in (\ref{eq:angular}).
\label{prop:normal}
\end{proposition}

\color{black}

For both densities (\ref{eq:Langevin}) and (\ref{eq:angular}), the modes can be explicitly determined from $M$. Let the spectral decomposition of $M$ be $M = U \Lambda U^T$, where $U$ is an $m \times m$ orthogonal matrix and $\Lambda = \text{diag}(\lambda_1, \ldots, \lambda_m)$ with $\lambda_1 \geq \cdots \geq \lambda_m$. The following proposition characterizes the modes. The proof is provided in the supplementary material.

\begin{proposition}
The mode of (\ref{eq:Langevin}) or (\ref{eq:angular}) is given by $P_0 = U_1 U_1^T$, where $U_1$ contains the first $k$ columns of $U$. The mode is unique if and only if $\text{rank}(M) \geq k$ and $\lambda_k > \lambda_{k+1}$.
\label{prop:mode}
\end{proposition}


\textcolor{revision}{
This proposition implies that when (\ref{eq:Langevin}) or (\ref{eq:angular}) is used as a prior distribution, the dominant eigenvectors of $M$ can be chosen to encode prior knowledge about $P$, and the magnitudes of the associated eigenvalues indicate the strength of this prior belief. For example, setting $M = 0$ yields the uniform distribution on $\mathcal{P}_{m, k}$, corresponding to a non-informative prior. A vague prior can be specified by taking $M = s P_{0}$ with a small positive $s$ and an initial guess $P_{0}$.
If we believe that a particular direction $\gamma$ lies in the subspace spanned by $P$, we may set
$M = a \gamma \gamma^T + s I_m$ for $a > 0$,
where larger values of $a$ represent stronger confidence in this prior information. 
For multiple directions, one may use
$M = a \gamma_1 \gamma_1^T + b \gamma_2 \gamma_2^T + s I_m$
with $a$ and $b$ tuned to reflect the relative strength of confidence in each direction. The effects of these choices of $M$ are further illustrated by simulation in Section~\ref{subsection:prior}.
}


\section{Model Specification}
\label{section:model}

This section presents the model specification, recommends prior distributions, and derives the corresponding posterior distributions.

Let the observations be $\{(y_{ij}, x_{ij}), y_{ij}\in\mathbb{R}^{r}, x_{ij}\in\mathbb{R}^p, i = 1, \ldots, n, j = 1, \ldots, J_i\}$,
where $(y_{ij}, x_{ij})$ denotes the data collected from subject $i$ at time point $j$. 
Let
$Y_{i} = (y_{i1}, \ldots, y_{iJ_{i}}) \in\mathbb{R}^{r\times J_{i}}$ and
$X_{i} = (x_{i1}, \ldots, x_{iJ_{i}}) \in\mathbb{R}^{p\times J_i}$
be all responses and covariates for the $i$th subject.
Consider the following linear regression model with latent variables,
for $i = 1, \ldots, n$,
\begin{align}
	Y_{i} = \Gamma Z_{i} + \Gamma_{0}Z_{0, i}, \quad
	Z_{i} = \mu 1_{J_{i}}^{T} + \eta X_{i} + \zeta_{i}, \quad
	Z_{0, i} = \mu_{0} 1_{J_{i}}^{T} + \zeta_{0, i},
\label{eq:latent}
\end{align}
where $(\Gamma, \Gamma_{0})$ is an $r\times r$ orthogonal matrix,
$\Gamma \in \mathbb{R}^{r\times u}$,
$\Gamma_{0} \in \mathbb{R}^{r\times (r-u)}$,
$Z_{i}\in\mathbb{R}^{u\times J_i}$,
$Z_{0, i} \in\mathbb{R}^{(r-u)\times J_i}$,
$\mu \in\mathbb{R}^{u}$,
$\eta\in\mathbb{R}^{u\times p}$,
$\zeta_{i} \in\mathbb{R}^{u\times J_i}$,
$\mu_{0}\in\mathbb{R}^{r-u}$, and
$\zeta_{0, i}\in\mathbb{R}^{(r-u)\times J_i}$.
The model (\ref{eq:latent}) explicitly writes the response $Y_i$ as the sum of two components, where $Z_{i}$ depends on the covariate $X_{i}$ whereas $Z_{0, i}$ does not.
Following the envelope-model framework, we assume that $\zeta_i$ and $\zeta_{0, i}$ are uncorrelated. Specifically, we assume that they jointly follow a scalar mixture of matrix-variate normal distributions as follows.
\begin{align}
	\binom{\zeta_{i}}{\zeta_{0, i}}
	\sim \text{SMMN}(0_{r\times J_i}, \text{diag}\{\Omega, \Omega_{0}\}, R_{i}, G),
\label{eq:xi}
\end{align}
where $\Omega \in \mathbb{R}^{u\times u}$ and
$\Omega_{0} \in \mathbb{R}^{(r-u)\times (r-u)}$ are positive definite, and
$R_i = R_{i}(\rho) \in\mathbb{R}^{J_i\times J_i}$ is a correlation matrix parameterized by $\rho$.
It follows that the marginal distribution of $Y_i$ is
\begin{align}
  Y_{i}
  \sim \text{SMMN}(\alpha 1_{J_i}^T + \beta X_i, \Sigma_{\varepsilon}, R_i, G),
\label{eq:model-Y}
\end{align}
where
\begin{align}
	\alpha = \Gamma \mu + \Gamma_{0} \mu_{0} \in\mathbb{R}^r, \quad
    \beta = \Gamma\eta \in\mathbb{R}^{r\times p},
    \quad
    \Sigma_{\varepsilon}
    = \Gamma \Omega \Gamma^T + \Gamma_0 \Omega_0 \Gamma_0^T
    \in\mathbb{R}^{r\times r}.
\label{eq:model-env}
\end{align}
We further assume that observations from different subjects are independent,
that is, $Y_i$ and $Y_{i'}$ are independent for $i\neq i'$.
The latent model (\ref{eq:latent})-(\ref{eq:xi}), or equivalently the marginal model (\ref{eq:model-Y})-(\ref{eq:model-env}), are referred to as the {\bf Ro}bust {\bf L}ongitudinal {\bf E}nvelope {\bf M}odel (RoLEM).
As a special case, when the distribution in (\ref{eq:xi}) or (\ref{eq:model-Y}) is normal, the model reduces to the {\bf L}ongitudinal {\bf E}nvelope {\bf M}odel (LEM).

The parameter $\Sigma_{\varepsilon} $ quantifies the variance and covariance of different responses at each time point.
For any subject $i$, each column of $Y_i$, for example, $\text{col}_j(Y_{i})$, contains all responses  at time point $j$. Its variance is
\begin{align*}
  var(\text{col}_j(Y_{i})) = c \Sigma_{\varepsilon},
\end{align*}
where $c = E(\tau^{-1})$ for $\tau \sim G(\cdot ; \nu)$ is a constant only depending on $G$.
\textcolor{revision}{Thus, the variance of the multivariate outcome is the same at every time point.}
Similarly, each row of $Y_i$, for example, $\text{row}_k(Y_{i})$, contains the $k$th response across all time points for the $i$th subject. Its variance is
\begin{align*}
  var(\text{row}_k(Y_{i})) = c \sigma_{kk} R_i,
\end{align*}
where $\sigma_{kk}$ is the $(k, k)$th entry of $\Sigma_{\varepsilon}$.
Here, $R_i$ accounts for the temporal correlation of repeated measurements of the same response through a parameter $\rho$.
\textcolor{revision}{It implies that all responses share the same temporal correlation structure.}
Common choices for $R_{i}$ include compound symmetry (CS),
where $R_i = (1 - \rho)I_{J_i} + \rho 1_{J_i}1_{J_i}^T$,
or AR(1), where $R_i = (\rho^{|s - l|})$.

To ensure identifiability, we impose constraints on $\{\Gamma, \Gamma_{0}, \Omega, \Omega_{0}\}$. Recall that $\Gamma$ is an orthogonal matrix, and the projection matrix $P = \Gamma\Gamma^{T}$ is identifiable as long as the linear space spanned by $\Gamma$ remains unchanged. Because the values of $\Omega$ and $\Omega_{0}$ depend on the particular choice of $\Gamma$ and $\Gamma_{0}$, it suffices to specify a unique and deterministic procedure to compute $\Gamma$ from a given $P$.
Let $(U_{1}, U_{2})$ be any chosen orthogonal matrix, where $U_{1}\in\mathbb{R}^{r\times u}$ and $U_{2}\in\mathbb{R}^{r\times (r-u)}$.
For any given $P \in \mathcal{P}_{r, u}$, compute $\tilde\Gamma \in\mathbb{R}^{r\times u}$, whose columns are the first $u$ eigenvectors of $P$. Because the first $u$ eigenvalues of $P$ are all ones, $\tilde\Gamma$ may not be unique. Next, let $A = (U_{2}^{T}\tilde\Gamma)(U_{1}^{T}\tilde\Gamma)^{-1} \in \mathbb{R}^{(r-u)\times u}$, which is uniquely determined whenever $U_{1}^{T}\tilde\Gamma$ is invertible.
Finally, compute
\begin{align}
    \Gamma = (U_{1} + U_{2} A)(I_{u} + A^{T}A)^{-1/2}, \quad
	\Gamma_0 = (-U_1A^T + U_2)(I_{r-u} + AA^T)^{-1/2}.
\label{eq:A2Gamma}
\end{align}
This construction parameterizes $P$ through the matrix $A$, which reduces to the method of Chakraborty and Su \cite{chakraborty2023comprehensive} when $U_{1} = (I_{u}, 0)^{T}$ and $U_{2} = (0, I_{r-u})^{T}$. This general form provides greater flexibility, particularly when $(I_{u}, 0)^{T} \tilde \Gamma$ is not invertible.

\subsection{Prior distributions}

This subsection describes the prior distributions for the model parameters.
For ease of presentation, we focus on the case where the SMMN follows a matrix-variate 
$t$-distribution; that is, $G(\cdot; \nu)$ is taken to be Gamma($\nu/2$, $\nu/2$), so that 
$\nu$ represents the degrees of freedom.
The full parameter vector is
\begin{align*}
   \theta
   = \{\alpha, \beta, \Sigma_{\varepsilon}, \rho, \nu\}
   = \{\alpha, \eta, P, \Omega, \Omega_{0}, \rho, \nu\}.
\end{align*}
The dimension $u$ and the correlation structure are assumed to be known.
Model selection for these components will be addressed in Section~\ref{subsection:selection}.

\begin{itemize}
\item 
The prior distribution of the intercept $\alpha \in \mathbb{R}^r$ is non-informative, that is,
$$\pi(\alpha)\propto 1.$$
It is the same prior specification used in Khare et al \cite{khare2017bayesian} and Chakraborty and Su \cite{chakraborty2023comprehensive}.

\item 
The prior distribution of $\eta\in\mathbb{R}^{u\times p}$, conditional on $\Gamma$ and $\Omega$, is a matrix-variate normal distribution $\text{MN}(\Gamma^T\xi, \Omega, H^{-1})$,
where $\xi\in\mathbb{R}^{r\times p}$ and $H\in\mathbb{R}^{p\times p}$ are user-specified matrices.
\begin{align*}
  \pi(\eta\mid \Gamma, \Omega)
  \propto
  |\Omega|^{-p/2}
  \text{etr}\left(-\frac{1}{2}\Omega^{-1}(\eta - \Gamma^T\xi)H(\eta - \Gamma^T\xi)^T\right).
\end{align*}
This prior is conjugate and is also used in Khare et al \cite{khare2017bayesian} and Chakraborty and Su \cite{chakraborty2023comprehensive}.
The prior is vague when $H$ is close to zero and becomes non-informative when $H = 0$.

\item 
The matrices $\Omega \in\mathbb{R}^{u\times u}$ and $\Omega_0\in\mathbb{R}^{(r-u)\times (r-u)}$ are both positive definite.
Their prior distributions are independent inverse-Wishart distributions with different parameters.
\begin{align*}
  \pi(\Omega)
  &\propto
  |\Omega|^{-(k + u + 1)/2}
  \text{etr}\left(-\frac{1}{2}\Psi \Omega^{-1}\right), \\
  \pi(\Omega_0)
  &\propto
  |\Omega_0|^{-(k_0 + r - u + 1)/2} \text{etr}\left(-\frac{1}{2}\Psi_0\Omega_0^{-1} \right),
\end{align*}
where the degrees of freedom $k > 0$ and $k_{0} > 0$, and scale parameters $\Psi\in\mathbb{R}^{u\times u}$ and $\Psi_0\in\mathbb{R}^{(r-u)\times (r-u)}$ are positive definite.

\item 
The prior distribution of $P \in \mathcal{P}_{r, u}$  
is a matrix Langevin distribution with parameter $M\in\mathbb{R}^{r\times r}$, where $M$ is symmetric.
\begin{align*}
  \pi(P) \propto \text{etr}(MP).
\end{align*}
Setting $M = 0$ yields a non-informative prior, 
whereas choosing $M$ to be close to zero leads to a vague prior. 
Prior knowledge can also be incorporated through an appropriate choice of $M$,
as discussed in Section~\ref{subsection:Grassmann}.

\item 
We consider two correlation structures, CS and AR(1), in this paper.
Let $\rho \in (0, 1)$ and the prior distribution is uniform($0$, $1$).
\begin{align*}
  \pi(\rho) \propto 1.
\end{align*}

\item 
The degrees of freedom $\nu$ are positive. 
Assume $\nu > 2$ to ensure the existence of the second moment of $Y$. 
The prior distribution of $\nu$ is Gamma($a$, $b$) with $a > 0$ and $b > 0$.
\begin{align*}
  \pi(\nu) \propto  \nu^{a-1} e^{-b\nu}.
\end{align*}
There exists a large literature on choosing a prior for $\nu$ in $t$-regression 
\cite{fernandez1999multivariate,
fonseca2008objective,
wang2012bayesian,
ding2014bayesian}.
The choice of Gamma distribution is a compromise between simplicity and flexibility.
\end{itemize} 

\textcolor{revision}{
The parameters in the prior distributions are referred to as hyperparameters. They are specified by the user to reflect prior knowledge about $\theta$, the model parameter. Section~\ref{subsection:hyper-pars} provides a discussion on the selection of hyperparameters.}

\subsection{Posteriors}
\label{subsection:posterior}

The density of the joint posterior distribution is
\begin{align*}
  \pi(\theta\mid Y, X)
  \propto
  \pi(Y\mid X, \theta)
  \pi(\alpha)\pi(\eta\mid\Gamma, \Omega)\pi(\Omega)\pi(\Omega_0)\pi(P)\pi(\rho)\pi(\nu).
\end{align*}
It is difficult to work with this posterior distribution directly.
We introduce a latent variable $\tau = \{\tau_1, \ldots, \tau_n\}$ and consider the hierarchical model
\begin{align*}
    Y_i\mid \tau_i
    &\sim \text{MN}(\alpha 1_{J_i}^T + \beta X_i, \tau_i^{-1} \Sigma_{\varepsilon}, R_i), \\
    \tau_i
    &\sim \text{Gamma} \left(\frac{\nu}{2}, \frac{\nu}{2}\right).
\end{align*}
\textcolor{revision}{The latent variable $\tau_i$ can be incorporated into either $\Sigma_{\varepsilon}$ or $R_i$, both of which yield the same marginal distribution. We choose to include $\tau_i$ in $\Sigma_{\varepsilon}$ in order to keep $R_i$ as a correlation matrix.}
When $\tau_i = 1$, $Y_i$ follows a normal distribution. A large value of $\tau_i$ indicates that $Y_i$ is a potential outlier.
The density of the joint posterior for the augmented data is
\begin{align*}
  \pi(\theta, \tau\mid Y, X)
  \propto
  \pi(Y\mid X, \theta, \tau) \pi(\tau)
  \pi(\alpha)\pi(\eta\mid\Gamma, \Omega)\pi(\Omega)\pi(\Omega_0)\pi(P)\pi(\rho)\pi(\nu).
\end{align*}
We derive the conditional posterior distribution of each parameter given the remaining ones, $Y$, and $X$ as follows. Some of them are standard distributions, while others are not and require an implementation of the Metropolis-Hastings algorithm.

Let
\begin{align*}
	\Delta_{i}
	= \text{tr}\big(
   \Sigma_{\varepsilon}^{-1} (Y_i - \alpha 1_{J_i}^T - \beta X_i)
   R_i^{-1} (Y_i - \alpha 1_{J_i}^T - \beta X_i)^T\big).
\end{align*}

\begin{itemize}
\item
The conditional posterior of $\tau_i$ is a
Gamma($(\nu + J_{i}r)/2$, $(\nu + \Delta_{i})/2$).

\item
The conditional posterior of $\alpha$ is a multivariate normal distribution $N(\mu_{\alpha}, \Sigma_{\alpha})$, where
\begin{align*}
    \mu_{\alpha}
    &= \left(\sum_{i=1}^n  \tau_i(1_{J_i}^TR_i^{-1}1_{J_i}) \right)^{-1} \sum_{i=1}^n \tau_i[(Y_i - \beta X_i)R_i^{-1}1_{J_i}], \\
    \Sigma_{\alpha}
    &= \left(\sum_{i=1}^n  \tau_i(1_{J_i}^TR_i^{-1}1_{J_i}) \right)^{-1} \Sigma_{\varepsilon}.
\end{align*}

\item
The conditional posterior of $\eta$ is a matrix-variate normal distribution $\text{MN}(\Gamma^{T}\tilde\xi, \Omega, \tilde H^{-1})$, where
\begin{align*}
    \tilde H
    &= H + \sum_{i=1}^n \tau_i X_iR_i^{-1}X_i^T, \\
    \tilde \xi
    &= \left(\xi H + \sum_{i=1}^n\tau_i(Y_i - \alpha 1_{J_i}^T)R_i^{-1}X_i^T\right)\tilde H^{-1}.
\end{align*}

\item
The conditional posterior of $\Omega$ is an inverse-Wishart distribution $IW(k + p + \sum_{i}J_{i}, \tilde\Psi)$,
where
\begin{align*}
    \tilde\Psi
    = \Psi + (\eta - \Gamma^T\xi)H(\eta - \Gamma^T\xi)^T
    + \sum_{i=1}^n \tau_i
  \Gamma^T(Y_i - \alpha 1_{J_i}^T - \beta X_i)R_i^{-1}
  (Y_i - \alpha 1_{J_i}^T - \beta X_i)^T
   \Gamma.
\end{align*}

\item
The conditional posterior of $\Omega_0$ is an inverse-Wishart distribution $IW(k_{0} + \sum_{i}J_{i}, \tilde\Psi_{0})$,
where
\begin{align*}
    \tilde\Psi_0
     = \Psi_0 + \sum_{i=1}^n \tau_i \Gamma_0^T
  (Y_i - \alpha 1_{J_i}^T)R_i^{-1}
  (Y_i - \alpha 1_{J_i}^T)^T
   \Gamma_0.
\end{align*}

\item
The conditional posterior of $P$ is not a standard distribution. Its density is
\begin{align*}
  \pi(P\mid\cdots)
  &\propto
  \exp\left(-\frac{1}{2}\sum_{i=1}^n \tau_i\Delta_i
  -\frac{1}{2}\text{tr}(\Omega^{-1}(\eta - \Gamma^T\xi)H(\eta - \Gamma^T\xi)^T)
  +
  \text{tr}(MP)\right),
\end{align*}
where $\cdots$ means the remaining parameters, $Y$, and $X$.

\item
The conditional posterior of $\rho$ is not a standard distribution. Its density is
\begin{align*}
  \pi(\rho\mid\cdots)
  &\propto
  \exp\left(-\frac{r}{2}\sum_{i=1}^n \log|R_i| -\frac{1}{2}\sum_{i=1}^n\tau_i \Delta_i
  \right).
\end{align*}

\item
The conditional posterior of $\nu$ is not a standard distribution. Its density is
\begin{align*}
  \pi(\nu\mid\cdots)
  &\propto
  \exp\left(
  \frac{n\nu}{2}\log\frac{\nu}{2}
  -n\log\Gamma\left(\frac{\nu}{2}\right) 
  + \left(\frac{\nu}{2}-1\right)\sum_{i=1}^n\log\tau_i -\frac{\nu}{2}\sum_{i=1}^n\tau_i
  + (a - 1) \log\nu - b\nu \right).
\end{align*}
\end{itemize}

\section{Implementation}
\label{section:implementation}

This section addresses practical aspects of implementation, including the Metropolis-Hastings algorithm, specification of hyper-parameters, selection of initial values, model selection, and software.

\subsection{Metropolis-Hastings algorithm}
\label{MH-algorithm}

Samples of $\{\theta, \tau\}$ are drawn from the posterior distribution in the order of
$\tau_i$, $\nu$, $\Omega$, $\Omega_0$, $P$, $\rho$, $\alpha$, and $\eta$.
We can compute $\Gamma$ and $\Gamma_0$ from $P$ by (\ref{eq:A2Gamma}) whenever is necessary.
The conditional posterior distributions of $\tau_i$, $\Omega$, $\Omega_0$, $\alpha$, and $\eta$ are standard distributions and hence they can be sampled directly. We implement Metropolis-Hastings algorithms \cite{gelman2013bayesian} to sample $\nu$, $P$, and $\rho$ from their conditional posteriors.

\begin{itemize}
\item 
When sampling $\rho$ with the current value being $\rho^{(s)}$, a proposed value $\rho^{*}$ is drawn from
$\text{uniform}(\rho^{(s)} - \delta_{\rho}, \rho^{(s)} + \delta_{\rho})$, where $\delta_{\rho} > 0$ is a pre-specified constant.
Because $\rho\in(0, 1)$, we adjust $\rho^{*}$ if it falls outside this interval:  
replace $\rho^{*}$ by $|\rho^{*}|$ if $\rho^{*} < 0$, and by $2 - \rho^{*}$ if $\rho^{*} > 1$.
Let $f(\rho\mid\rho^{(s)})$ be the density of the proposal distribution. 
This proposal distribution is symmetric \textcolor{revision}{in the sense that $f(\rho\mid\rho^{(s)}) = f(\rho^{(s)}\mid \rho)$}, and hence it yields a random walk update.

\item 
When sampling $\nu$ with the current value being $\nu^{(s)}$, a proposed value $\nu^*$ is drawn from
$\text{uniform}(\nu^{(s)} - \delta_{\nu}, \nu^{(s)} + \delta_{\nu})$, where $\delta_{\nu} > 0$ is a pre-specified constant.
Because $\nu > 2$, we adjust $\nu^{*}$ if it falls below 2 by replacing it by $4 - \nu^{*}$ if $\nu^{*} < 2$.
Let $f(\nu\mid\nu^{(s)})$ be the density of the proposal distribution. 
This proposal distribution is symmetric \textcolor{revision}{in the sense that $f(\nu\mid\nu^{(s)}) = f(\nu^{(s)}\mid \nu)$}, resulting in a random walk update.

\item 
When sampling $P$ with the current value $P^{(s)}$, a proposed matrix $P^*$ is drawn from the density $f(P\mid P^{(s)}) = |W|^{-u/2}|I_r - P + W^{-1}P|^{-r/2}$,
where {\color{revision} $W = \sigma^2 I_r + P^{(s)}$ and $\sigma^2$ is a pre-specified constant.}
This proposed matrix can be generated as $P^* = Z(Z^TZ)^{-1}Z^T$, where $Z$ is drawn from a matrix-variate normal distribution $\text{MN}(0, W, I)$.
{\color{revision} This proposal distribution is symmetric in the sense that $f(P\mid P^{(s)}) = f(P^{(s)} \mid P)$, which is proved in the supplementary material.
Hence, this step is also a random walk. 
}
\end{itemize}

The values of $\delta_{\rho}$, $\delta_{\nu}$, $\sigma^2$ control the acceptance rates of the Metropolis-Hastings algorithm. It is recommended that the values be chosen to ensure the acceptance rate falls within $(0.2, 0.5)$.
\textcolor{revision}{The convergence of the MCMC chains can be assessed using the standard diagnostic tools, such as trace plots, autocorrelation plots, and formal convergence tests.}

\subsection{Hyper-parameters in prior distributions}
\label{subsection:hyper-pars}

The hyperparameters of the prior distributions need to be specified. 
Users may choose them based on prior knowledge, following common practice in Bayesian inference. When limited prior information is available, we recommend the following settings for vague priors:
\begin{itemize}
\item 
set $\xi = 0_{u\times p}$ and $H = 10^{-3} I_p$ for the prior of $\eta$;
\item 
set $k = u + 1$ and $\Psi = 10^{-3} I_u$ for the prior of $\Omega$;
\item 
set $k_0 = r-u + 1$ and $\Psi_0 = 10^{-3} I_{r-u}$ for the prior of $\Omega_0$; and 
\item 
set $M = 10^{-3} I_r$ for the prior of $P$.
\end{itemize}
In some cases, even weaker prior information may be desired, in which case one may instead set 
$H = 10^{-6} I_p$, $\Psi = 10^{-6} I_u$, $\Psi_0 = 10^{-6} I_{r - u}$, and $M = 10^{-6} I_r$ to obtain nearly non-informative priors.
Finally, for the Gamma-prior for the degrees of freedom $\nu$, set $a = 1.4$ and $b = 0.04$,
which makes the prior probability of $P(2 < \nu < 110) = 0.95$.

\subsection{Initialization}

Under mild conditions, an MCMC chain starting from any value will converge if the chain is long enough.
In practice, to ensure faster convergence, we recommend setting the initial value of the chain as a raw estimate of the parameters, assuming the distribution of $Y_i$ is normal.
Notice that if the distribution of $Y_i$ is normal, the mean and variance of $Y_i$ are
\begin{align*}
  \mu_i &= \alpha 1_{J_i}^T + \beta X_i \in \mathbb{R}^{r\times J_i}, &
  \Lambda_i &= R_i\otimes \Sigma_{\varepsilon} \in \mathbb{R}^{J_ir \times J_ir}.
\end{align*}
Hence, the initial values can be obtained by the steps below.

\begin{enumerate}
\item
Set $\tau_i^{(0)} = 1$ for all $i = 1, \ldots, n$.
Set $\nu^{(0)} = 10$.

\item
Set $\alpha^{(0)}$ and $\beta^{(0)}$ as the least squares estimates of the intercepts and slopes when fitting individual regression with each component of $Y_i$ as response and $X_i$ as covariates.

\item
Set $\Sigma_{\varepsilon}^{(0)}$ as the variance matrix of all $r_{ij}$, 
where residual $r_{ij} = y_{ij} - \alpha^{(0)} - \beta^{(0)} x_{ij}$.

\color{revision}

\item 
Set $\Gamma^{(0)}$ as the matrix formed by the first $u$ left singular vectors
from the singular value decomposition of $\beta^{(0)}$.

\item 
Compute $A^{(0)}$ from $\Gamma^{(0)}$ and then compute $\Gamma_{0}^{(0)}$ from $A^{(0)}$ as in (\ref{eq:A2Gamma}).

\item 
Set $\eta^{(0)}= {\Gamma^{(0)}}^{T}\beta^{(0)}$,
$\Omega^{(0)} = {\Gamma^{(0)}}^{T} \Sigma_{\varepsilon}^{(0)} \Gamma^{(0)}$,  
and $\Omega_{0}^{(0)} = {\Gamma_{0}^{(0)}}^{T} \Sigma_{\varepsilon}^{(0)} \Gamma_{0}^{(0)}$.

\item 
Set $\rho^{(0)}$ as the correlation between $\tilde r_{ijk}$ and $\tilde r_{i, j+1, k}$,
where $\tilde r_{ijk} = r_{ijk} / \sqrt{\sigma_{kk}}$,
$r_{ijk}$ is the $k$th component of $r_{ij}$,
and $\sigma_{kk}$ is the $(k, k)$th entry of  $\Sigma_{\varepsilon}^{(0)}$.
\end{enumerate}
\color{revision}
Note that the last step is valid because, for both the CS and AR(1) correlation structures, the correlation between $r_{ijk}$ and $r_{i,j+1,k}$ equals $\rho$ for each $k$th component of residuals.

We also need to choose an orthogonal matrix $U$, which is used in (\ref{eq:A2Gamma}) to compute $\Gamma$ and $\Gamma_0$ from $A$, or to compute $A$ from $P$. This matrix remains fixed throughout the sampling procedure. 
A simple choice is $U = I_r$ or a randomly generated orthogonal matrix.
If the trace plot of $A$ exhibits many extreme values, it may indicate that the current choice of $U$ is not appropriate, as it leads to inverting a near-singular matrix. In such cases, select a different $U$ and restart the sampling procedure. 
The following two strategies usually lead to a good choice.
\begin{itemize}
\item 
Set $U$ as the $Q$-matrix from the QR factorization of an initial estimate of $\beta$, for example, the least squares estimate as in Step 2 above. If $r > p$, append additional columns to make $U$ a full orthogonal matrix.

\item 
Set $U = (\hat{\Gamma}, \hat{\Gamma}_0)$,
where $\hat{\Gamma}$ and $\hat{\Gamma}_0$ are the posterior means of $\Gamma$ and $\Gamma_0$ from the initial sampling using $U = I_r$ or other choice of matrix. 
\end{itemize}.

\subsection{Model selection}
\label{subsection:selection}

When applying RoLEM, it is necessary to determine the envelope dimension $u$, the correlation structure, and whether to assume $t$-distributed or normally distributed random errors. These decisions can all be formulated as a model selection problem. There is a rich literature on this topic in Bayesian analysis; in this paper, we focus on BIC and WAIC.

The BIC (Bayesian or Schwarz information criterion) \cite{schwarz1978estimating} is defined as
\begin{align*}
   \text{BIC} = -2\log  L + p_{\text{BIC}}\log n,
\end{align*}
where 
$L$ is the maximized likelihood, 
$p_{\text{BIC}}$ is the effective number of parameters, 
and $n$ is the number of subjects. 
In practice, $L$ can be approximated by the maximum of the likelihood observed across MCMC draws.
For RoLEM, 
\begin{align*}
  p_{\text{BIC}} = r + up + (r-u)u + u(u+1)/2 + (r-u)(r-u+1)/2 + 1 + 1, 
\end{align*}
which corresponds to $\alpha$, $\eta$, $P$, $\Omega$, $\Omega_0$, $\rho$, and $\nu$. 
BIC achieves a trade-off between goodness-of-fit and model complexity, 
with a smaller value indicating a better model.  
Comparing two models using BIC is asymptotically equivalent to computing the Bayes factor \cite{kass1995bayes}.  
Specifically, the Bayes factor $\text{BF}_{12}$ of two models $M_1$ and $M_2$ is approximated by 
$\text{BF}_{12} \approx \exp( -(\text{BIC}_1 - \text{BIC}_2)/2)$, 
where $\text{BIC}_1$ and $\text{BIC}_2$ are BIC values of $M_1$ and $M_2$, respectively. 
A difference of $\text{BIC}_1 - \text{BIC}_2 \leq -6$ corresponds to a Bayes factor  larger than 20, and provides strong evidence in favor of $M_1$ over $M_{2}$ \cite{kass1995bayes}.

Out-of-sample prediction performance is commonly used for model comparison. 
Although cross-validation can estimate out-of-sample prediction, it is often time-consuming.  
The WAIC (widely applicable or Watanabe-Akaike information criterion) \cite{gelman2013bayesian, watanabe2010asymptotic} is shown to be asymptotically equivalent to cross-validation. It is defined by 
\begin{align*}
   \text{WAIC} 
   = -2\sum_{i=1}^{n} \log E_{\text{post}}[p(y_{i}\mid\theta)]
    + 2\sum_{i=1}^{n} var_{\text{post}}(\log p(y_{i}\mid\theta)),
\end{align*}
where $E_{\text{post}}$ and $var_{\text{post}}$ denote the expectation and variance  with respect to the posterior distribution of $\theta$.
The first term represents the log pointwise predictive density, while
the second term corresponds to the effective number of parameters.
In practice, these quantities can be approximated by their sample counterparts using posterior samples. 
Models with smaller WAIC values are preferred.

In practice, BIC and WAIC can be computed for different candidate models, which may differ in terms of the envelope dimension $u$, the correlation structure, or both. The model with the smallest BIC or WAIC is preferred. 



%

\subsection{Software}

The algorithm has been implemented in an R package, with the core computational components written in \verb"C++". 
The package is available at \verb"https://github.com/pzengauburn/benvlp". 
Table~\ref{table:time} reports the average computation time (in seconds) required to generate 1,000 samples for selected scenarios. 
All computations were performed on a MacBook Pro with an Apple M2 chip and 8 GB of RAM, running macOS Tahoe 26.1. 
An AR(1) correlation structure was assumed in these computations; results are similar for other correlation structures. 
The computation time increases as $r$, $p$, $n$, or $J$ grows. 
Although the total number of observations is the same for 
$(n, J) = (100, 5)$ and $(50, 10)$, as well as for $(n, J) = (200, 5)$ and $(100, 10)$, computation time increases as $J$ becomes larger.

\begin{table}[ht]
\centering
\caption{Average Computation Time (in seconds) Required to Generate 1,000 Samples}
\begin{tabular}{lccccccc}
\hline 
& & \multicolumn{3}{c}{$J = 5$} & \multicolumn{3}{c}{$J = 10$} \\
$(r, p)$ & $u$ & $n = 50$ & $n = 100$ & $n = 200$ & $n = 50$ & $n = 100$ & $n = 200$ \\
\hline 
$(5, 6)$ 
& $2$ & $0.14$ & $0.27$ & $0.52$ & $0.31$ & $0.61$ & $1.18$ \\
& $3$ & $0.15$ & $0.27$ & $0.52$ & $0.31$ & $0.61$ & $1.20$ \\
& $4$ & $0.15$ & $0.27$ & $0.53$ & $0.30$ & $0.60$ & $1.16$ \\
\hline 

$(20, 30)$ 
& $2$ & $0.89$ & $1.68$ & $3.28$ & $1.71$ & $3.35$ & $6.55$ \\
& $3$ & $0.89$ & $1.64$ & $3.28$ & $1.72$ & $3.36$ & $6.57$ \\
& $4$ & $0.89$ & $1.70$ & $3.29$ & $1.73$ & $3.35$ & $6.57$ \\
\hline 
\end{tabular}
\label{table:time}
\end{table}

In both the simulation studies and the real data analysis, we use equal time points for each subject for simplicity and clarity of presentation. However, the model can accommodate the imbalanced case. Recall that for each subject, the response $Y_i \in \mathbb{R}^{r \times J_i}$ contains $J_i$ time points, and $J_i$ need not be the same across subjects. The implementation provided in the associated R package can also handle data with unequal time points.

\color{black}

\section{Examples}
\label{section:example}

This section discusses some numerical experiments and a real data analysis to demonstrate the performance of RoLEM.

\subsection{Simulation study}
\label{subsection:performance}

This subsection reports simulation studies to compare the performance of RoLEM with several alternative methods under various settings. 

Follow the steps below to generate synthetic datasets.
Set
the number of responses to \textcolor{revision}{$r = 20$},
the number of covariates to \textcolor{revision}{$p = 30$}, and
the number of time points to $J = 5$ or 10.
Set the envelope dimension to $u = 3$.
Randomly generate a matrix $A\in\mathbb{R}^{(r-u)\times u}$, whose entries are independently sampled from uniform$(-1, 1)$. 
Let $U = I_r$ and compute $\Gamma$ and $\Gamma_0$ using (\ref{eq:A2Gamma}). 
Randomly generate $\Omega$ and $\Omega_0$ as diagonal matrices whose diagonal entries are independently drawn from uniform$(0, 1)$ and uniform$(5, 10)$, respectively. Finally, compute
$
   \Sigma_{\varepsilon}
   = \Gamma \Omega \Gamma^{T} + \Gamma_{0} \Omega_{0} \Gamma_{0}^{T}$.
For each subject, generate the covariate matrix $X_{i} \in \mathbb{R}^{p\times J}$ with entries independently drawn from $N(0, 1)$.
Compute the response as $Y_{i} = \alpha 1^{T}_{J} + \Gamma\eta X_{i} + \varepsilon_{i}$,
where the entries of $\alpha$ and $\eta$ are independently sampled from uniform$(-5, 5)$ and
$\varepsilon_{i}$ is drawn from MT$(4, 0, \Sigma_{\varepsilon}, R(\rho))$ with an AR(1) correlation structure and $\rho = 0.5$.
The number of subjects is $n = 50$, 100, or 200.
\color{revision}
All parameters, covariates $X$, and response $Y$ are regenerated for each synthetic dataset. 

For each synthetic dataset, we apply three methods separately:
\begin{itemize}
\item 
RLMM: Robust Bayesian linear mixed-effects model assuming $t$-distributed random errors,
\item 
LEM: Longitudinal envelope model assuming normally distributed random errors, and
\item 
RoLEM: Robust longitudinal envelope model assuming $t$-distributed random errors. 
\end{itemize}
Note that RLMM does not include an envelope structure. 
We therefore compare the following scenarios in the simulation study:
\begin{itemize}
\item 
Number of subjects: $n = 50$, 100, 200.
\item 
Number of time points: $J = 5$, 10.
\item
Estimation method: RLMM, LEM, RoLEM.
\end{itemize}

\color{black}

Assume that the envelope dimension is known to be 3 and the correlation structure is known to be AR(1). 
Select the hyperparameters according to the nearly non-informative priors described in
Section~\ref{subsection:hyper-pars}.
Discard the first 10,000 burn-in samples, and then retain one out of every ten from the subsequent 100,000 samples. The estimates of the parameters are the posterior means.
We evaluate the performance of the estimates of $\beta$ and $\Sigma_{\varepsilon}$, which are denoted by $\hat\beta$ and $\hat\Sigma_{\varepsilon}$, using the Frobenius norm of their differences,
\begin{align*}
  D(\hat\beta, \beta) = \|\hat\beta - \beta\|_{F}, \quad
  D(\hat\Sigma_{\varepsilon}, \Sigma_{\varepsilon}) = \|\hat\Sigma_{\varepsilon} - \Sigma_{\varepsilon}\|_{F}, \quad
\end{align*}
where $\|M\|_{F} = \sqrt{\text{tr}(M^{T}M)}$ is the Frobenius norm of a matrix $M$.
\color{revision}
Additionally, we compute the 95\% highest posterior density (HPD) intervals for entries of $\beta$.

\begin{figure}[ht]
\centering
\includegraphics[width = 0.9\textwidth]{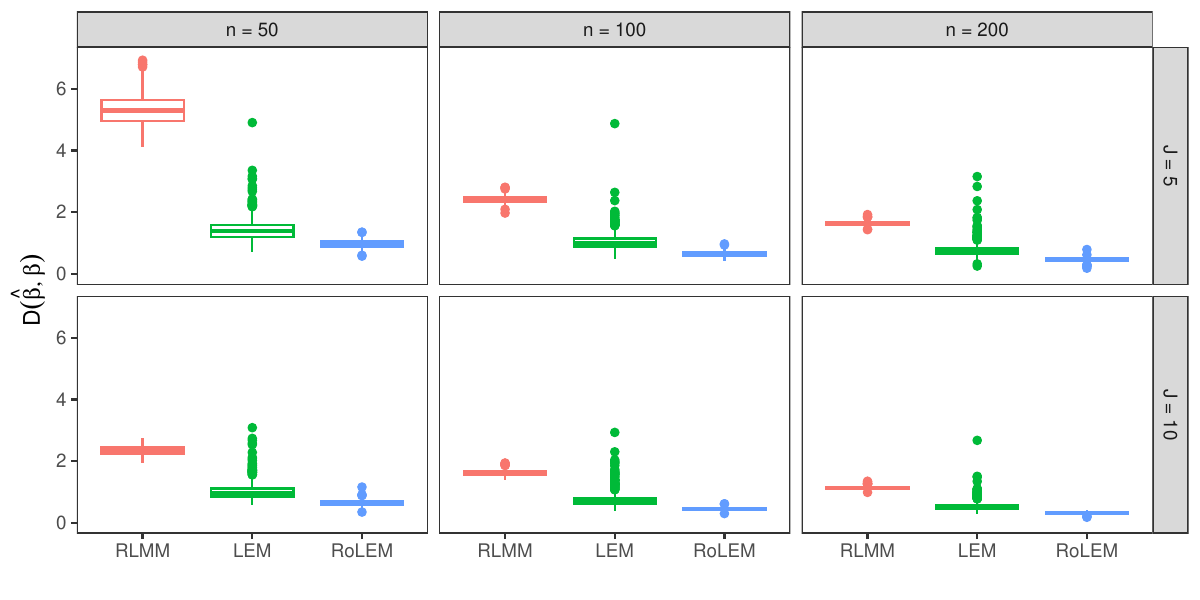}
\caption{Performance of estimating $\beta$ using RLMM, LEM, and RoLEM. Columns correspond to different sample sizes $n = 50$, 100, 200, and rows correspond to different time points $J = 5$ and $10$.}
\label{fig:ex1-t-beta}
\end{figure}

Based on 500 synthetic datasets, Figure~\ref{fig:ex1-t-beta} compares the performance of estimating $\beta$. 
Their numerical values are provided in the supplementary material.
Figure~\ref{fig:ex1-t-beta} shows that estimation accuracy improves as either the number of subjects or the number of time points increases, reflecting the increase in total sample size. 
RoLEM consistently outperforms LEM because the true random errors follow a $t$-distribution, whereas LEM assumes normally distributed errors. 
Both RoLEM and LEM outperform RLMM, demonstrating that identification of the underlying envelope structure improves estimation efficiency. 
When $n = 50$ and $J = 5$, the performance of RLMM is substantially worse than in other scenarios, highlighting the importance of capturing the envelope structure, particularly for smaller sample sizes.

\begin{figure}[ht]
\centering
\includegraphics[width = 0.9\textwidth]{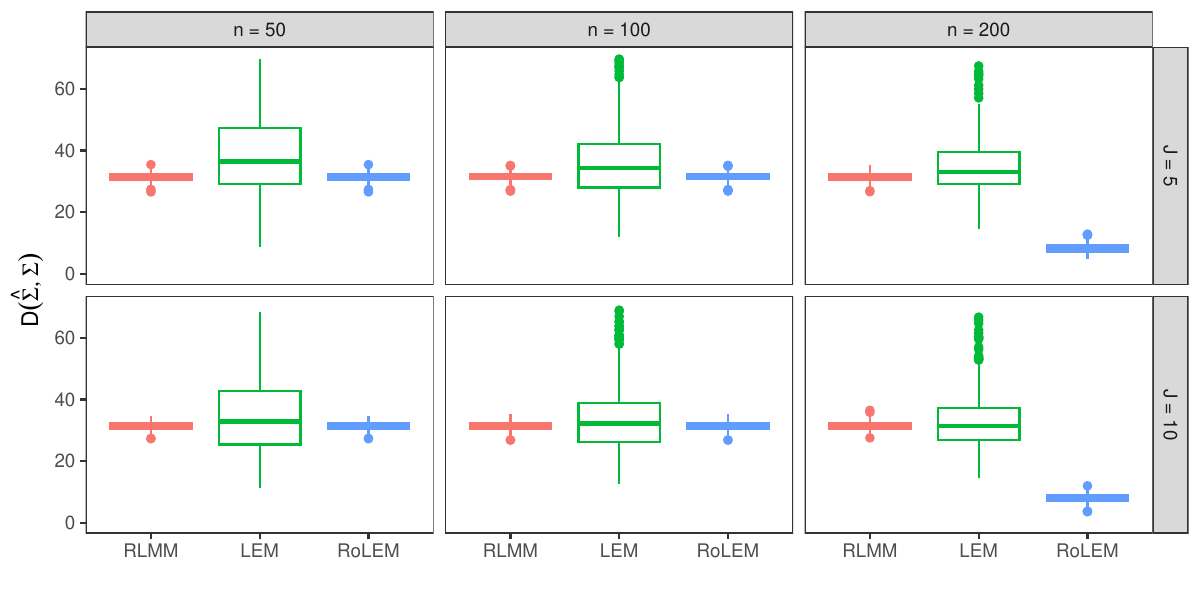}
\caption{Performance of estimating $\Sigma_{\varepsilon}$ using RLMM, LEM, or RoLEM. Columns correspond to different sample sizes $n = 50$, 100, 200, and rows correspond to different time points $J = 5$ and $10$.}
\label{fig:ex1-t-Sigma}
\end{figure}

Figure~\ref{fig:ex1-t-Sigma} compares the performance of estimating $\Sigma_{\varepsilon}$.
LEM exhibits much larger variability than RLMM and RoLEM because it assumes normal random errors, whereas the synthetic datasets are generated with $t$-distributed random errors.
The performances of RLMM and RoLEM are comparable in most scenarios, except when $n = 200$, where RoLEM performs better due to the efficiency gained from exploiting the envelope structure.

\begin{figure}[ht]
\centering
\includegraphics[width = 0.9\textwidth]{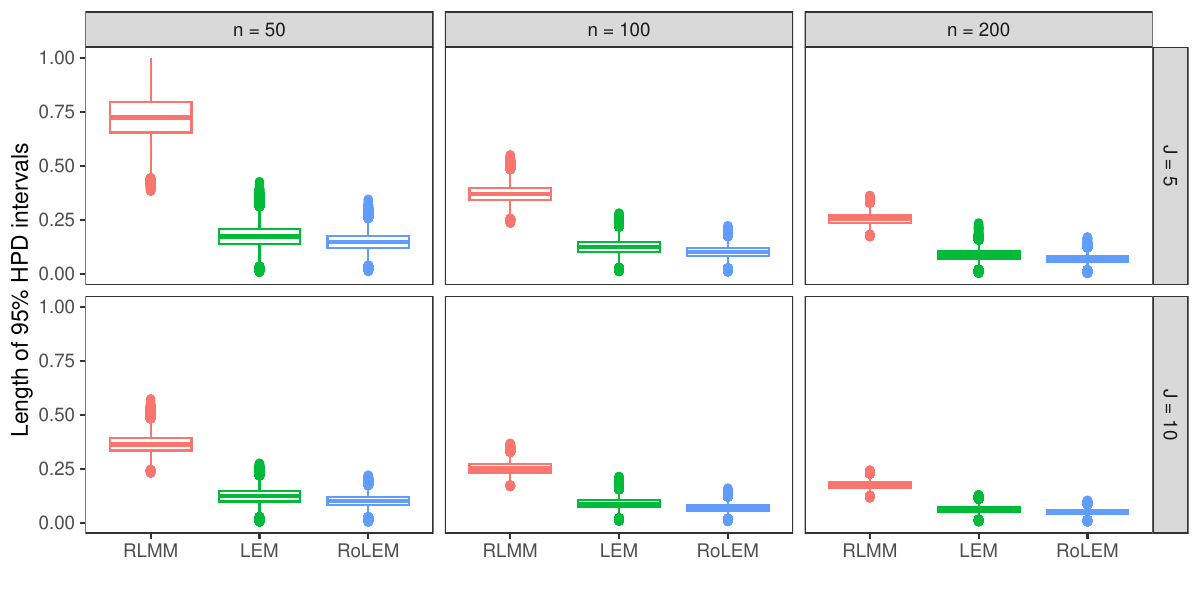}
\caption{\color{revision} Length of the 95\% HPD intervals for $\beta$ estimated by RLMM, LEM, and RoLEM. Columns correspond to different sample sizes $n = 50$, 100, 200, and rows correspond to different time points $J = 5$ and $10$.}
\label{fig:ex1-HPD-t-length}
\end{figure}

Figure~\ref{fig:ex1-HPD-t-length} displays boxplots of the lengths of the 95\% HPD intervals for $\beta$.
Since $\beta$ has $rp$ components, each boxplot contains all $rp$ components across 500 replicates.
The results show that RoLEM generally yields slightly shorter HPD intervals than LEM, and both have substantially shorter intervals than RLMM.
This highlights the advantage of the envelope structure, as it effectively removes variability in the responses that is unrelated to the covariates once the envelope is correctly identified.

\begin{figure}[ht]
\centering
\includegraphics[width = 0.9\textwidth]{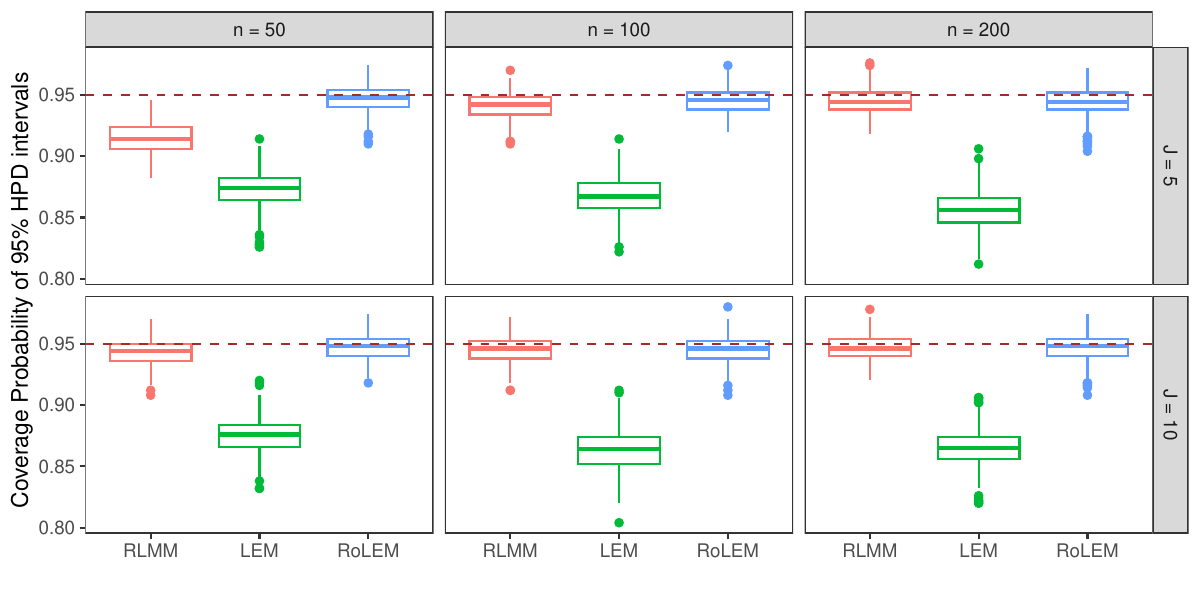}
\caption{\color{revision} Empirical coverage probabilities of the 95\% HPD intervals for $\beta$ using RLMM, LEM, and RoLEM. Columns correspond to different sample sizes $n = 50$, 100, 200, and rows correspond to different time points $J = 5$ and $10$.}
\label{fig:ex1-HPD-t-prob}
\end{figure}

We then compare the empirical coverage probabilities of the HPD intervals.
For each component of $\beta$, the empirical coverage probability is computed as the proportion of replicates in which the true value falls within the corresponding 95\% HPD interval.
Figure~\ref{fig:ex1-HPD-t-prob} displays the empirical coverage probabilities for all components of $\beta$. Ideally, these probabilities should be close to the nominal value of 0.95.
The empirical coverage of LEM is consistently below 0.95, due to its assumption of normal random errors. RLMM shows slightly lower coverage when $n = 50$ and $J = 5$ but aligns with the nominal level as the sample size increases. RoLEM performs as expected, with empirical coverages closely matching the nominal value.

In summary, RoLEM outperforms LEM because it accounts for $t$-distributed random errors, and it outperforms RLMM because leveraging the envelope structure enhances estimation efficiency.
The strength of envelope models lies not in improving the estimation of $\Sigma_{\varepsilon}$, but in identifying the component of variability in responses that is relevant to $\beta$. 
This, in turn, leads to improved estimation of $\beta$ and substantially shorter HPD intervals.

\color{black}

\subsection{Robustness}
\label{subsection:robust}

To evaluate the robustness of the proposed methods, we follow the data generation procedure in Section~\ref{subsection:performance}, but consider alternative ways to generate random errors.
Set $\varepsilon_{i} = L\epsilon_{i}B \in \mathbb{R}^{r \times J}$, where $\Sigma_{\varepsilon} = LL^{T}$ and $R(\rho) = BB^{T}$ and
\begin{itemize}
\item
Normal: the components of $\epsilon_{i}$ are iid $N(0, 2)$. In this case, $\varepsilon_{i} \sim \text{MT}(0, 2\Sigma_{\varepsilon}, R(\rho))$.
\item
Mixture normal: the components of $\epsilon_{i}$ are iid from a mixture-normal distribution $0.9 N(0, 1) + 0.1 N(0, 11)$.
\end{itemize}
Both the normal distribution and mixture-normal distribution are selected to have the same covariance matrix as the $t$-distribution used in Section~\ref{subsection:performance}.

\begin{figure}[ht]
\centering
\includegraphics[width = 0.9\textwidth]{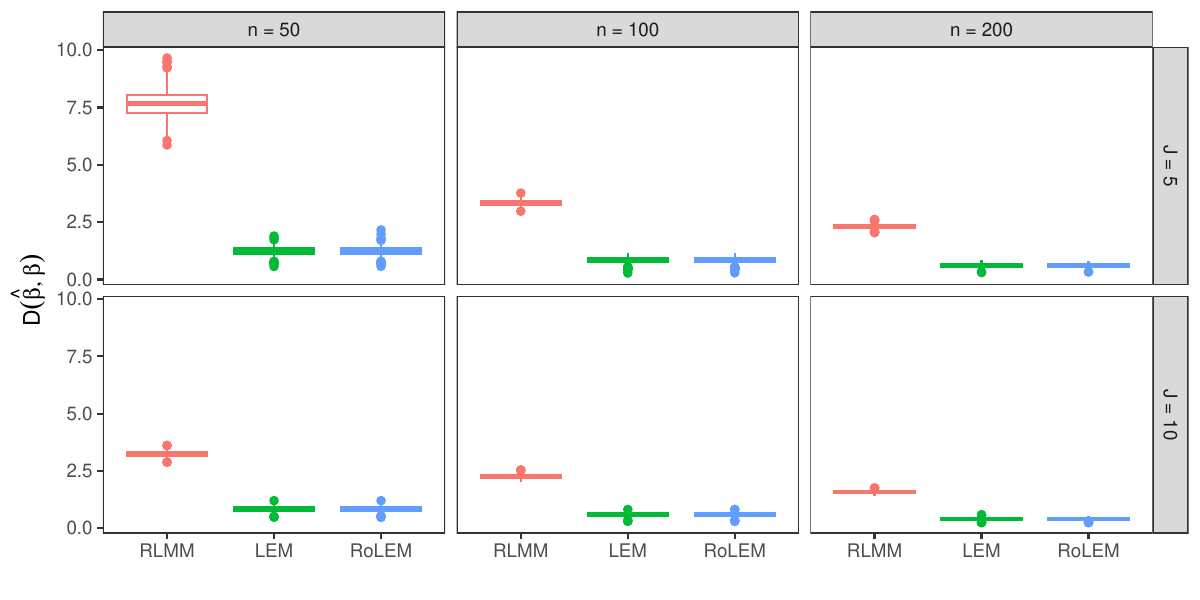}
\caption{Performance of estimating $\beta$ using RLMM, LEM, and RoLEM when the random errors follow a normal distribution. Columns correspond to different sample sizes $n = 50$, 100, 200, and rows correspond to different time points $J = 5$ and $10$.}
\label{fig:ex2-norm-beta}
\end{figure}

\begin{figure}[ht]
\centering
\includegraphics[width = 0.9\textwidth]{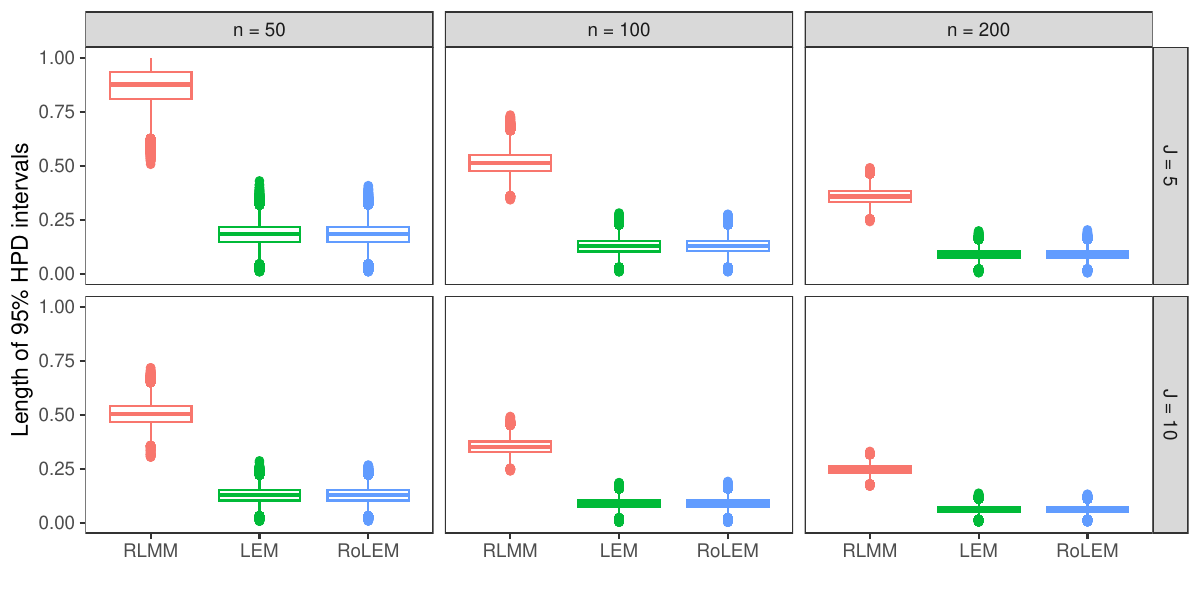}
\caption{Length of the 95\% HPD intervals for $\beta$ estimated by RLMM, LEM, and RoLEM when the random errors follow a normal distribution. Columns correspond to different sample sizes $n = 50$, 100, 200, and rows correspond to different time points $J = 5$ and $10$.}
\label{fig:ex2-HPD-norm-length}
\end{figure}

\begin{figure}[ht]
\centering
\includegraphics[width = 0.9\textwidth]{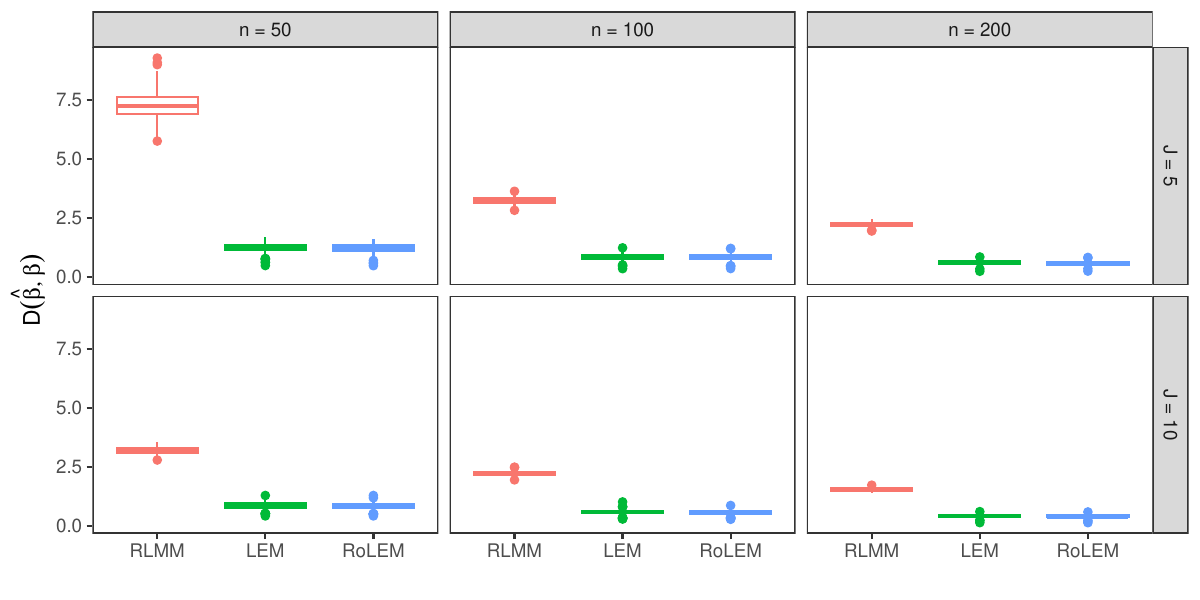}
\caption{Performance of estimating $\beta$ using RLMM, LEM, and RoLEM when the random errors follow a mixture-normal distribution. Columns correspond to different sample sizes $n = 50$, 100, 200, and rows correspond to different time points $J = 5$ and $10$.}
\label{fig:ex2-mix-beta}
\end{figure}

\begin{figure}[ht]
\centering
\includegraphics[width = 0.9\textwidth]{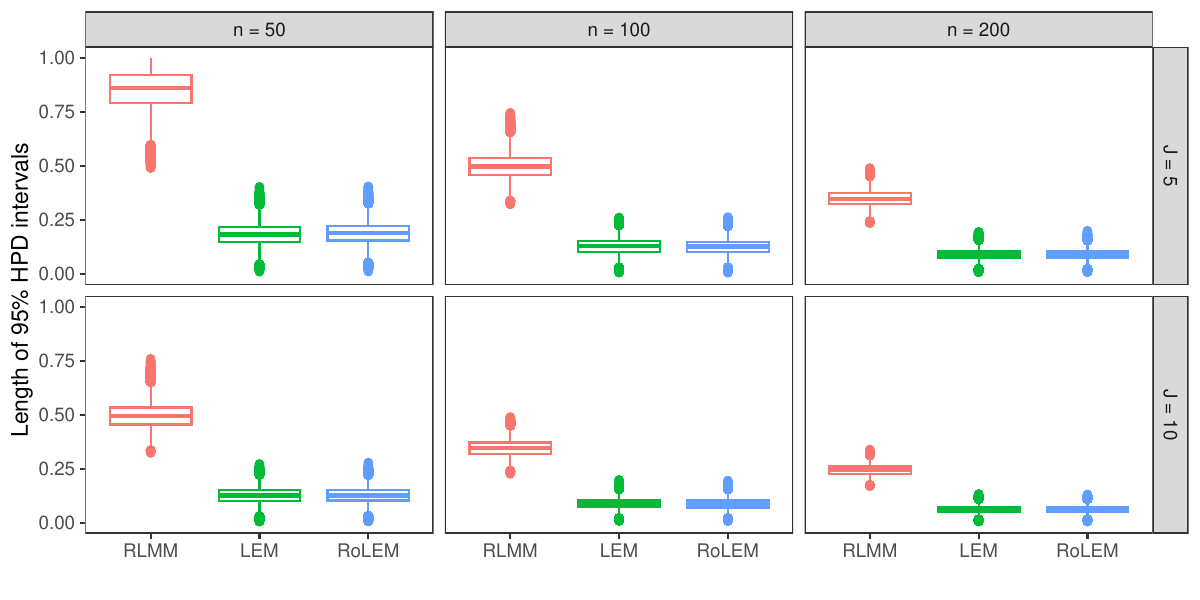}
\caption{Length of the 95\% HPD intervals for $\beta$ estimated by RLMM, LEM, and RoLEM when the random errors follow a mixture-normal distribution. Columns correspond to different sample sizes $n = 50$, 100, 200, and rows correspond to different time points $J = 5$ and $10$.}
\label{fig:ex2-HPD-mix-length}
\end{figure}

Repeat the simulation 500 times and evaluate the performance of estimating $\beta$ and $\Sigma_{\varepsilon}$, as in Section~\ref{subsection:performance}.
Figures~\ref{fig:ex2-norm-beta} and \ref{fig:ex2-HPD-norm-length} correspond to scenarios with normally distributed random errors, 
while Figures~\ref{fig:ex2-mix-beta} and \ref{fig:ex2-HPD-mix-length} correspond to scenarios with mixture-normal random errors.
\color{revision}
The numerical results are provided in the supplementary material.
Because the performance of estimating $\Sigma_{\varepsilon}$ and the coverage probabilities of the HPD intervals exhibit similar patterns, they are provided in the supplementary material to save space.

When the random errors follow a normal distribution, 
Figure~\ref{fig:ex2-norm-beta} shows that RoLEM and LEM perform comparably in estimating $\beta$, since the $t$-distribution includes the normal distribution as a special case. 
Both RoLEM and LEM still outperform RLMM, highlighting the advantage of leveraging the envelope structure to improve estimation efficiency. 
RLMM has slightly better performance than RoLEM and LEM in estimating $\Sigma_{\varepsilon}$, although this advantage decreases as the sample size increases. 
Nevertheless, Figure~\ref{fig:ex2-HPD-norm-length} shows that RLMM consistently produces much longer HPD intervals than RoLEM and LEM across all scenarios.
The empirical coverage probabilities of the HPD intervals for all three methods are generally close to the nominal value of 0.95, with slight deviations observed when $n = 50$ and $J = 5$.

When the random errors follow a mixture-normal distribution, 
Figure~\ref{fig:ex2-mix-beta} shows that RoLEM and LEM perform comparably in estimating $\beta$, and both outperform RLMM, highlighting the advantage of identifying the envelope structure. 
RoLEM slightly outperforms LEM and RLMM in estimating $\Sigma_{\varepsilon}$. 
Figure~\ref{fig:ex2-HPD-mix-length} shows that RoLEM and LEM provide much shorter HPD intervals than RLMM. 
The empirical coverage probabilities of the HPD intervals for all three methods are generally close to the nominal value of 0.95, with slight deviations observed when $n = 50$.

In summary, RoLEM demonstrates strong overall performance. 
When the random errors follow a normal distribution, it automatically adapts and achieves performance comparable to LEM. 
Even when the random errors follow a mixture-normal distribution, RoLEM maintains its robust performance. 
By identifying the envelope structure, RoLEM provides substantially shorter HPD intervals while preserving empirical coverage probabilities compared to standard methods.

\subsection{Model selection}

This subsection illustrates the performance of model selection using BIC and WAIC.
We generate synthetic datasets as described in Section~\ref{subsection:performance}.
The true envelope dimension is $u = 3$, and the correlation structure is AR(1).

We first assume that the correlation structure is known to be AR(1) and focus on selecting the envelope dimension $u$.
For each synthetic dataset, we fit models with different values of $u$ and compute the corresponding BIC and WAIC. The model with the smallest information criterion is selected.
The left part of Table~\ref{table:selection} reports the proportion of times each $u$ is selected across 500 replicates. Both BIC and WAIC correctly identify $u = 3$ in most scenarios.

Next, we assume that the envelope dimension is known to be $u = 3$ and focus on selecting the correlation structure from UNCOR, CS, and AR(1), where UNCOR denotes the uncorrelated case or $\rho = 0$. 
The right part of Table~\ref{table:selection} shows the proportion of times each correlation structure is selected across 500 replicates.
Both BIC and WAIC correctly identify AR(1) in all scenarios.

\begin{table}[htb]
\centering
\caption{Proportion of Selected Models Using BIC and WAIC}
\label{table:selection}
\begin{tabular}{llcccccc||ccc}
\hline
$(n, J)$ & Criterion & 1 & 2 & 3 & 4 & 5 & $6+$ & UNCOR & CS & AR(1) \\
\hline
$(50, 5)$ 
& BIC   & 0 & 0 & 0.998 & 0.002 & 0 & 0 & 0 & 0 &  1.000 \\
& WAIC  & 0 & 0 & 0.996 & 0.004 & 0 & 0 & 0 & 0 & 1.000 \\
\hline
$(50, 10)$ 
& BIC   & 0 & 0 & 1.000 & 0 & 0 & 0 & 0 & 0 &  1.000 \\
& WAIC  & 0 & 0 & 0.998 & 0.002 & 0 & 0 & 0 & 0 & 1.000 \\
\hline
$(100, 5)$ 
& BIC  & 0 & 0 & 1.000 & 0 & 0 & 0 & 0 & 0 & 1.000 \\
& WAIC & 0 & 0 & 1.000 & 0 & 0 & 0 & 0 & 0 &  1.000 \\
\hline
$(100, 10)$ 
& BIC  & 0 & 0 & 1.000 & 0 & 0 & 0 & 0 & 0 & 1.000 \\
& WAIC & 0 & 0 & 1.000 & 0 & 0 & 0 & 0 & 0 &  1.000 \\
\hline
$(200, 5)$ 
& BIC  & 0 & 0 & 1.000 & 0 & 0 & 0 & 0 & 0 & 1.000 \\
& WAIC & 0 & 0 & 1.000 & 0 & 0 & 0 & 0 & 0 &  1.000 \\
\hline
$(200, 10)$ 
& BIC  & 0 & 0 & 1.000 & 0 & 0 & 0 & 0 & 0 & 1.000 \\
& WAIC & 0 & 0 & 1.000 & 0 & 0 & 0 & 0 & 0 &  1.000 \\
\hline 
\end{tabular}
\end{table}

When both $u$ and the correlation structure are unknown, BIC and WAIC can be compared for models with each possible combination of $u$ and correlation structure. Although this approach is feasible, it requires substantially more computation. We do not explore it in the simulation studies but demonstrate it in the real data analysis.

\subsection{Prior distribution}
\label{subsection:prior}

This subsection demonstrates the impact of incorporating prior knowledge on the estimation of $\beta$. 
We modify the data generation procedure in Section~\ref{subsection:performance} slightly by fixing $A$ to be 
$A = (a, I_3, a, I_3, a, I_3, a, I_3, a)^T \in \mathbb{R}^{17\times 3}$, 
where $a = (-1, 1, 1)^T$. 
With this setup, the envelope is spanned by 
$\gamma_1 = 1_{5} \otimes (1, 1, 1, 1)^T/\sqrt{20}$, 
$\gamma_2 = 1_{5} \otimes (1, -1, 1, -1)^{T}/\sqrt{20}$, and 
$\gamma_3 = 1_{5} \otimes (1, 1, -1, -1)^{T}/\sqrt{20}$, 
which are three mutually orthogonal unit-length vectors in $\mathbb{R}^{20}$. 
Consider the following four different hyperparameters in the prior for $P$. 
\begin{align*}
M_1 &= s_0 I_{20}, &
M_2 &= s_1\gamma_1\gamma_1^T + s_0 I_{20}, \\
M_3 &= s_1(\gamma_1\gamma_1^T + \gamma_2\gamma_2^T) + s_0 I_{20}, &
M_4 &= s_1(\gamma_1\gamma_1^T + \gamma_2\gamma_2^T + \gamma_3\gamma_3^T) + s_0 I_{20}. 
\end{align*}
These four scenarios represent gradually increasing amounts of prior knowledge about $P$. 
In the simulation, we set $s_1 = 10^5$ and $s_0 = 10^{-6}$.

\begin{figure}[ht]
\centering
\includegraphics[width = 0.9\textwidth]{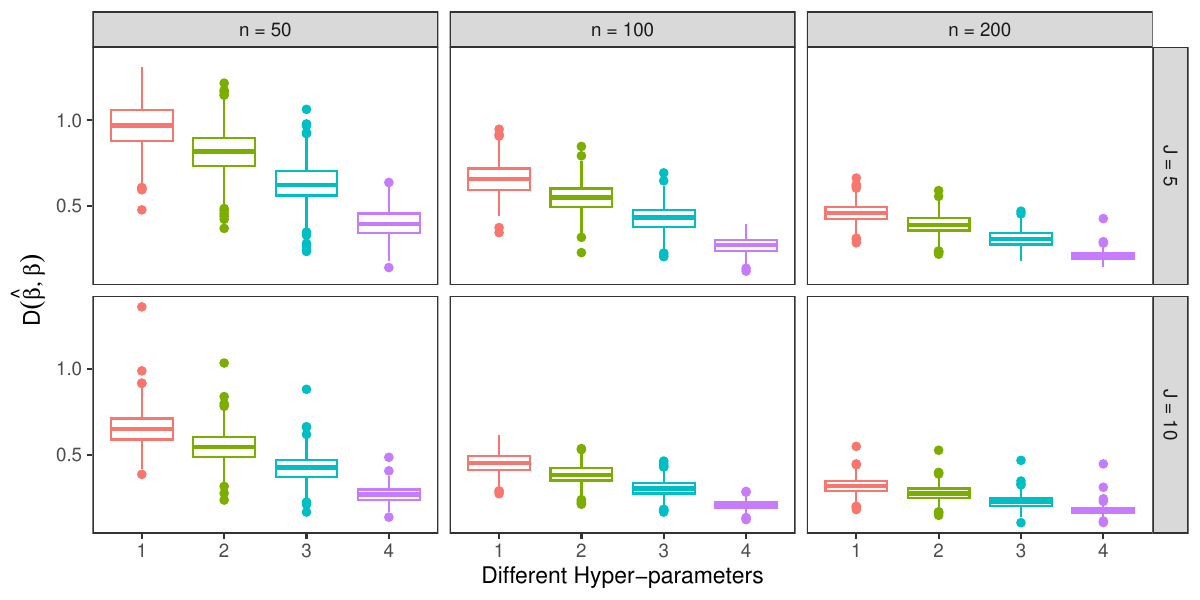}
\caption{\color{revision} Performance of estimating $\beta$ under different hyper-parameters in the prior for $P$. Columns correspond to different sample sizes $n = 50$, 100, 200, and rows correspond to different time points $J = 5$ and $10$.}
\label{fig:ex4-prior}
\end{figure}

Figure~\ref{fig:ex4-prior} displays the performance of estimating $\beta$ based on 500 replicates. As the prior distribution incorporates more information about $P$, the accuracy of estimating $\beta$ improves across all scenarios, demonstrating that prior knowledge has been effectively leveraged in the analysis.

\subsection{Real data analysis}

Existing literature has suggested that diabetes mellitus may be a potential risk factor for knee osteoarthritis \cite{Eitner2021}.
This study aims to evaluate the impact of diabetes mellitus on the physical and psychological health outcomes of patients with osteoarthritis in a longitudinal setting, accounting for potential confounding variables such as body mass index (BMI), osteoarthritis severity, and pain medication use.

The dataset is extracted from the Osteoarthritis Initiative (OAI) database 
(\verb"https://nda.nih.gov/oai"), 
which is a longitudinal cohort study of individuals with or at increased risk for knee osteoarthritis.
The impact of osteoarthritis on patients can be evaluated through both physical and mental health measures.
In this study, we focus on knee pain (KOOS and NRS), physical activity level (PASE), physical health status (PCS), and mental health status (MCS and CESD).
The primary covariate of interest is the presence of diabetes, 
and additional covariates include sex, age, BMI, pain medication use, and osteoarthritis severity (KL grade).
Descriptions of all variables are provided in Table~\ref{table:OAI-variables}.
After excluding observations with missing values, the final dataset consists of 221 osteoarthritis patients 
assessed at baseline and at 12-, 24-, 36-, and 48-month follow-ups.
Thus, the analysis involves $r=6$ response variables and $p=6$ covariates, with each patient contributing $J=5$ visits.

\begin{table}[ht]
\centering
\caption{List of Variables in the OAI Data}
\begin{tabular}{lll}
\hline
Variable & Description & Values \\
\hline
CESD & Center for Epidemiologic Studies Depression Scale & 0 -- 60 \\
KOOS & Knee Injury and Osteoarthritis Outcome Score & 0 -- 100 \\
MCS  & Mental Component Summary from SF-12 & 0 -- 100 \\
NRS  & Pain severity over past 30 days & 0 -- 10 \\
PASE & Physical Activity Scale for the Elderly &  0 -- 793 \\
PCS  & Physical Component Summary from SF-12 & 0 -- 100 \\
\hline
Age & Age in years  &  45 -- 83 \\
BMI & Body Mass Index & 18.50 -- 41.10 \\
Diabetes & Diabetes presence, 0 = No, 1 = Yes & 0, 1 \\
KL  & Kellgren-Lawrence grade for severity of osteoarthritis & 0 -- 4 \\
Medication & Pain Medication use, 0 = No, 1 = Yes & 0, 1 \\
Sex & Sex, 0 = Male, 1 = Female & 0, 1 \\
\hline
\end{tabular}
\label{table:OAI-variables}
\end{table}

\begin{figure}[ht]
\centering
\includegraphics[width = 0.3\textwidth]{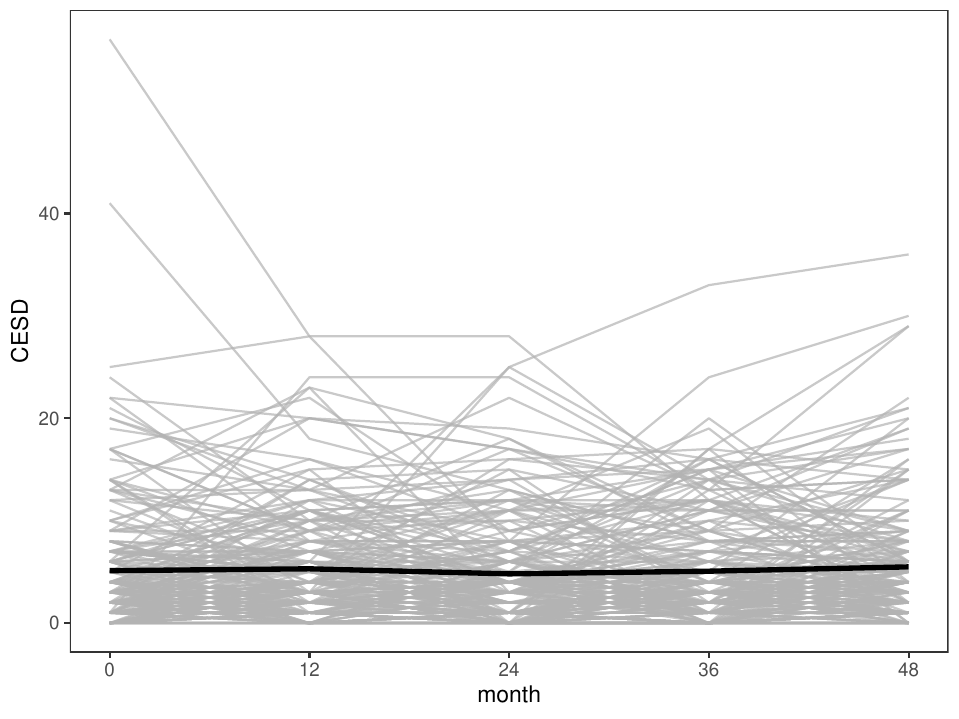}
\includegraphics[width = 0.3\textwidth]{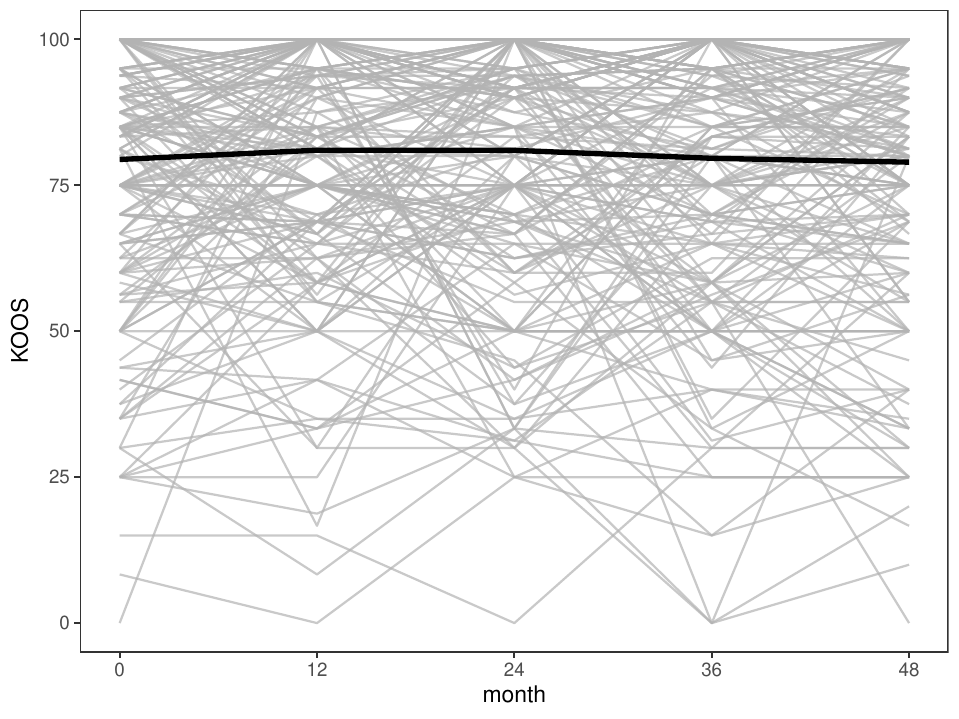}
\includegraphics[width = 0.3\textwidth]{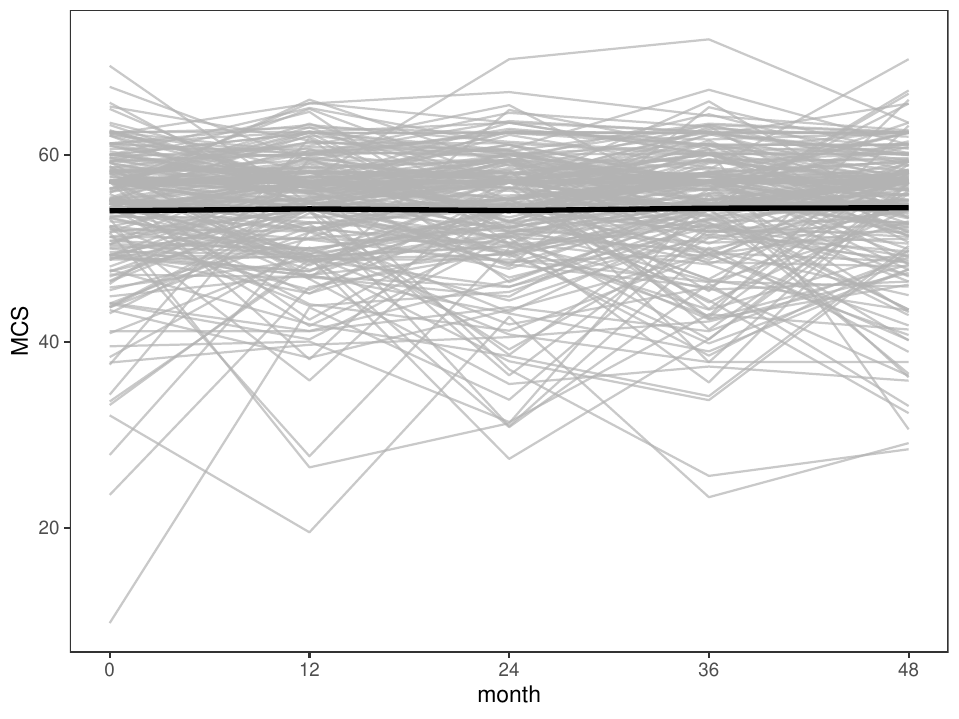}
\includegraphics[width = 0.3\textwidth]{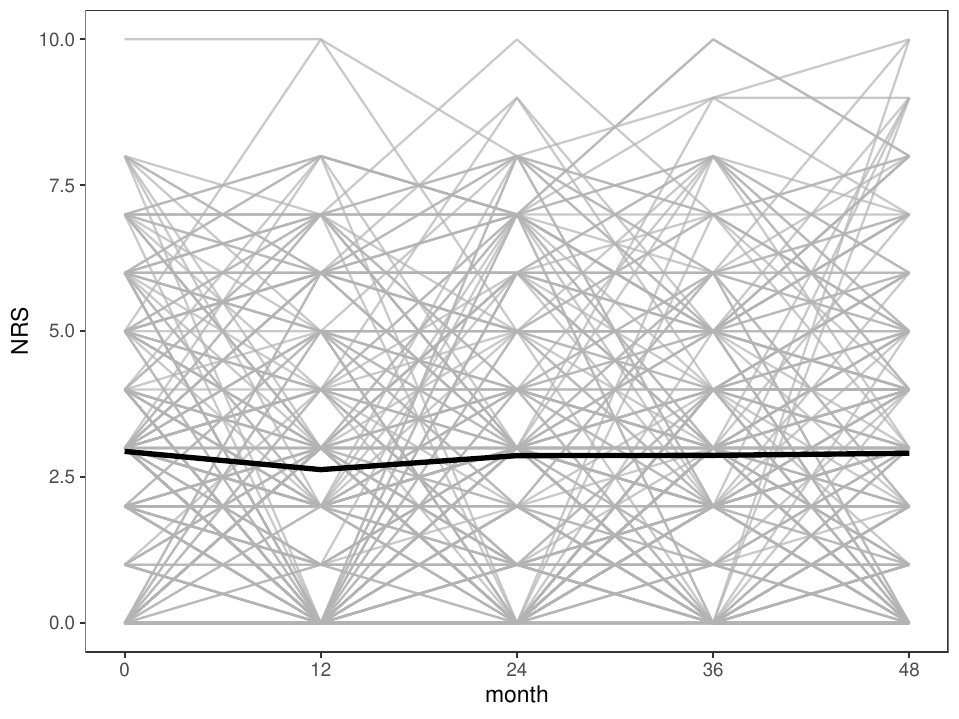}
\includegraphics[width = 0.3\textwidth]{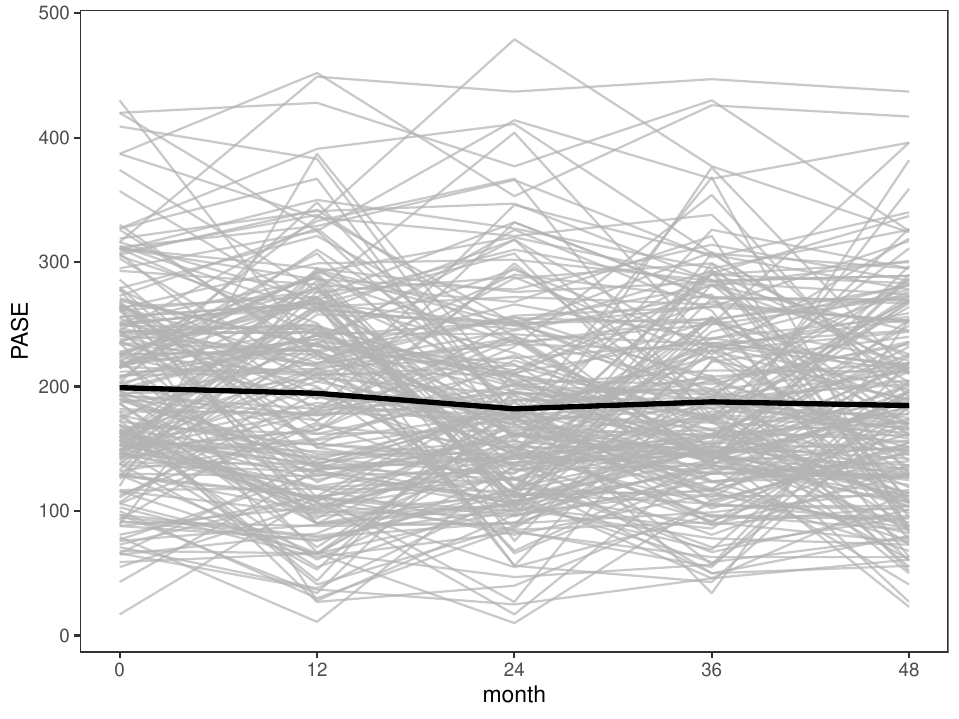}
\includegraphics[width = 0.3\textwidth]{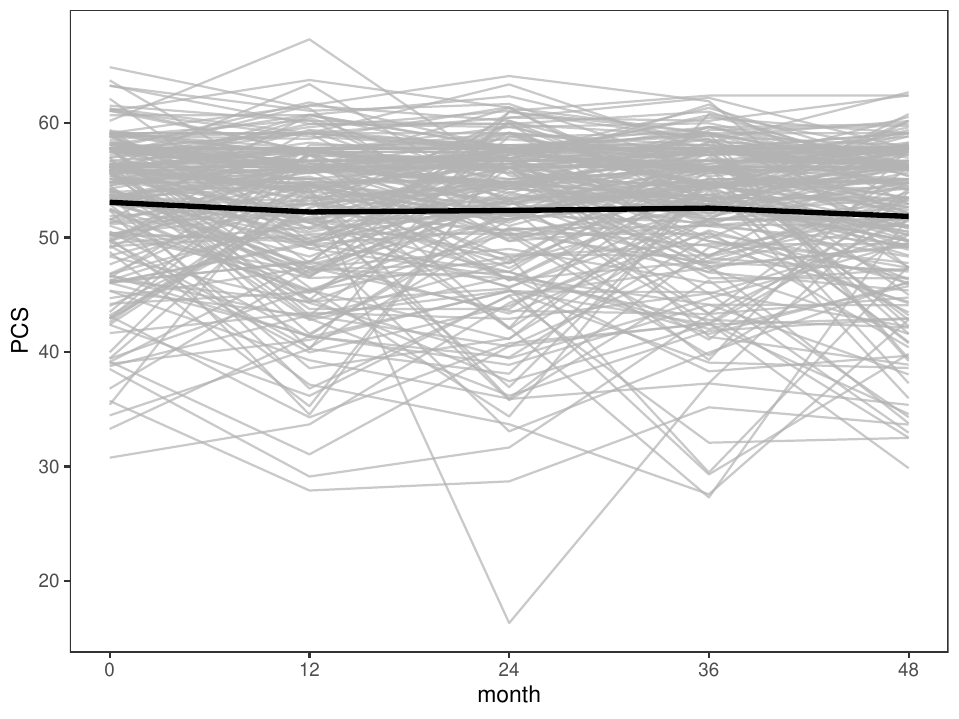}
\caption{The trajectories of each patient for each response variable, with the thick lines indicating the mean trajectories.}
\label{fig:OAI-y}
\end{figure}

Figure~\ref{fig:OAI-y} presents the trajectories of each patient for each response variable, with the thick lines indicating the mean trajectories. Although individual patient trajectories exhibit substantial variability, the mean trajectories remain relatively stable over time. Several individual trajectories also appear to be potential outliers.

All response and covariate variables were standardized to zero-mean and unit-variance before analysis. 
We applied the RoLEM to analyze the data, 
discarding the first 50,000 MCMC samples as burn-in and then generating 250,000 additional samples, 
retaining one out of every 50.
We compared WAIC and BIC values across models with different choices of $u$ and correlation structures in Table~\ref{table:OAI-BIC}.
Models with a compound-symmetric (CS) correlation structure consistently yielded smaller WAIC and BIC values than models with other correlation structures. 
The model with $u = 5$ attained the lowest WAIC overall, 
while the model with $u = 2$ achieved the smallest BIC. 
The WAIC approximates the out-of-sample prediction,
and the difference between the $u = 2$ and $u = 5$ models is roughly $1 - (14157.48/14167.93) = 0.07\%$. 
On the other hand, comparing BIC corresponds to evaluating the Bayes factor.
The difference of BIC values for the $u = 2$ model and the $u = 5$ model is large enough to claim very strong evidence in favor of the former. 
Hence, we proceed with the analysis using the model with $u=2$ and the CS correlation structure.  

\begin{table}[ht]
\centering
\caption{WAIC and BIC for Models with Different $u$ and Correlation Structure}

\begin{tabular}{llccccc}
\hline 
&& \multicolumn{5}{c}{Dimension} \\
 &
Correlation & 1 & 2 & 3 & 4 & 5 \\
\hline 
WAIC
& UNCOR & $16223.55$ & $16080.04$ & $16081.65$ & $16074.36$ & $16066.37$ \\
& AR(1) & $14657.11$ & $14600.87$ & $14599.85$ & $14599.18$ & $14584.64$ \\
& CS    & $14208.63$ & $14167.93$ & $14170.81$ & $14169.99$ & $14157.48$ \\

\hline 
BIC 
& UNCOR & $16284.09$ & $16165.80$ & $16184.14$ & $16191.24$ & $16198.57$ \\
& AR(1) & $14757.39$ & $14728.52$ & $14746.45$ & $14772.02$ & $14780.39$ \\
& CS    & $14319.96$ & $14301.85$ & $14325.45$ & $14351.61$ & $14362.29$ \\

\hline 
\end{tabular}
\label{table:OAI-BIC}
\end{table}

\begin{figure}[ht]
\centering
\includegraphics[width = 0.48\textwidth]{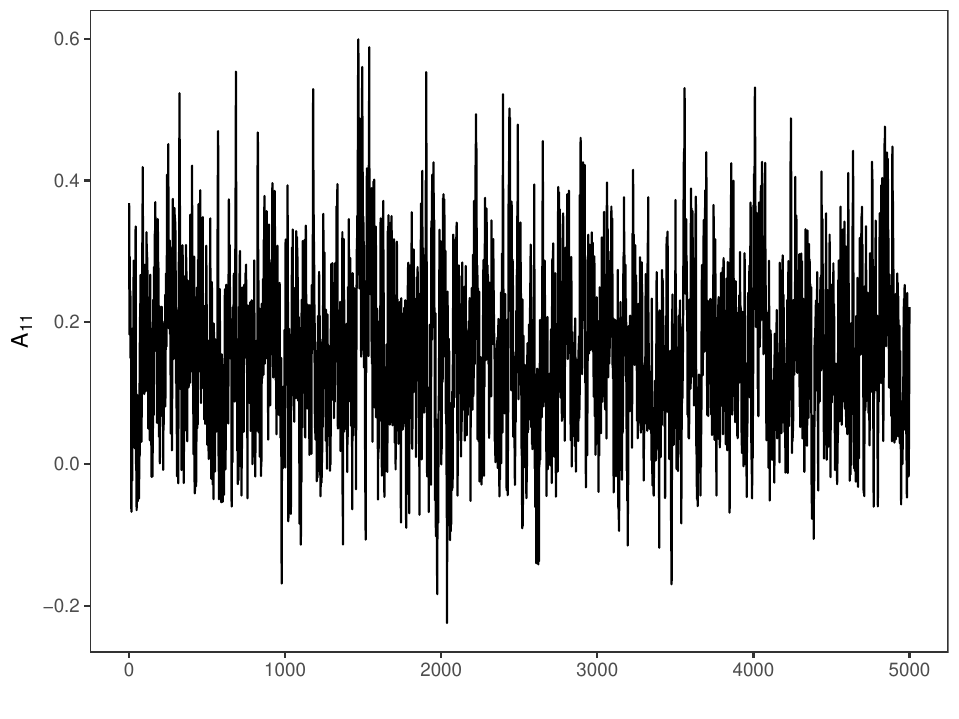}
\includegraphics[width = 0.48\textwidth]{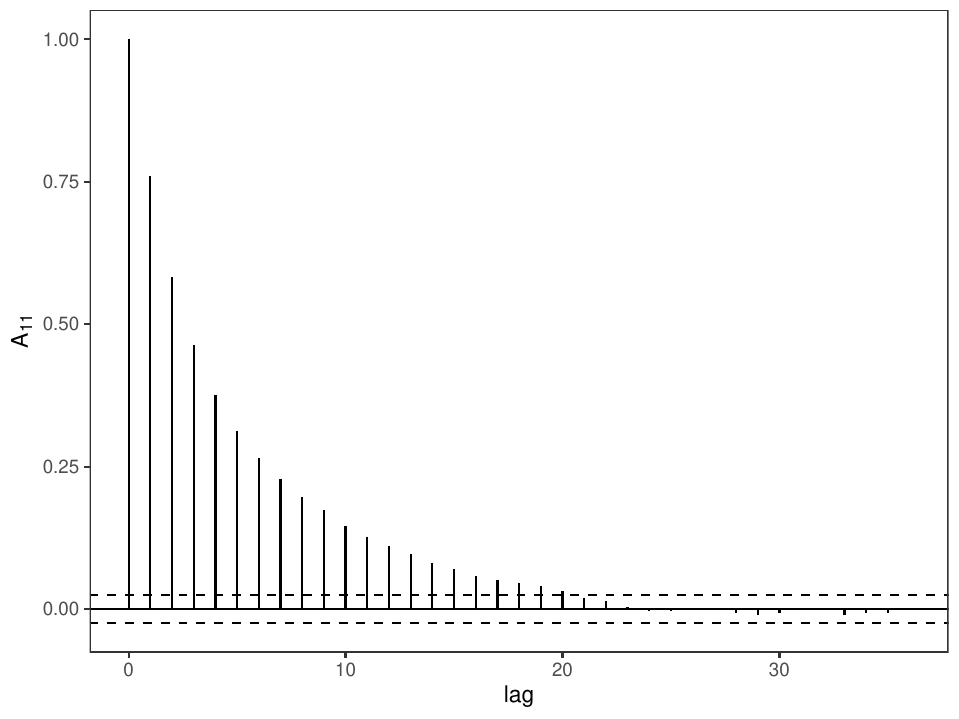}
\includegraphics[width = 0.48\textwidth]{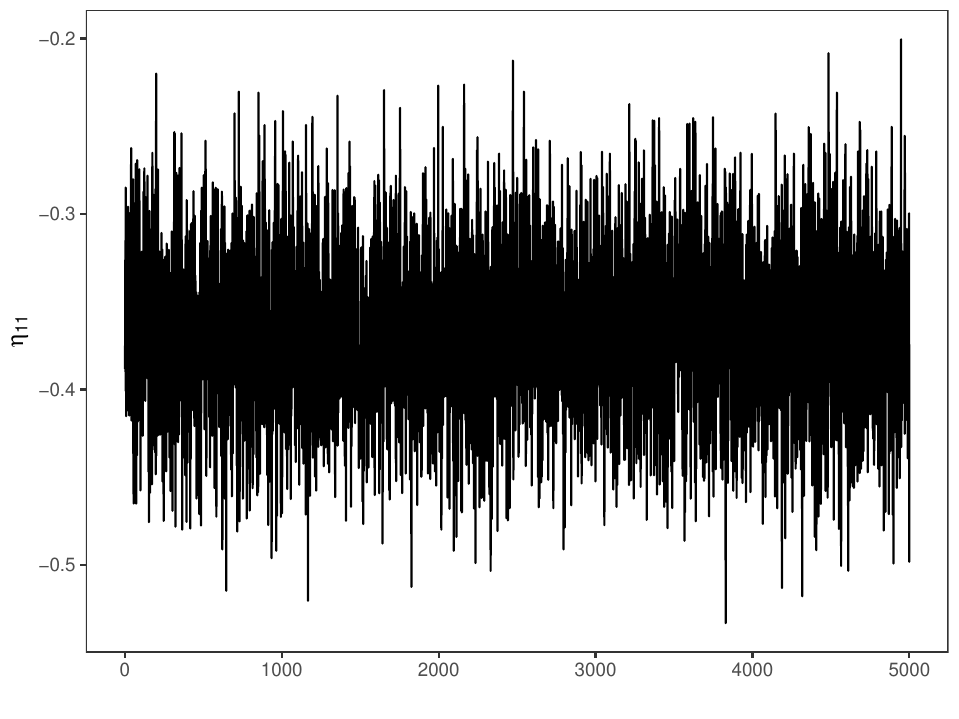}
\includegraphics[width = 0.48\textwidth]{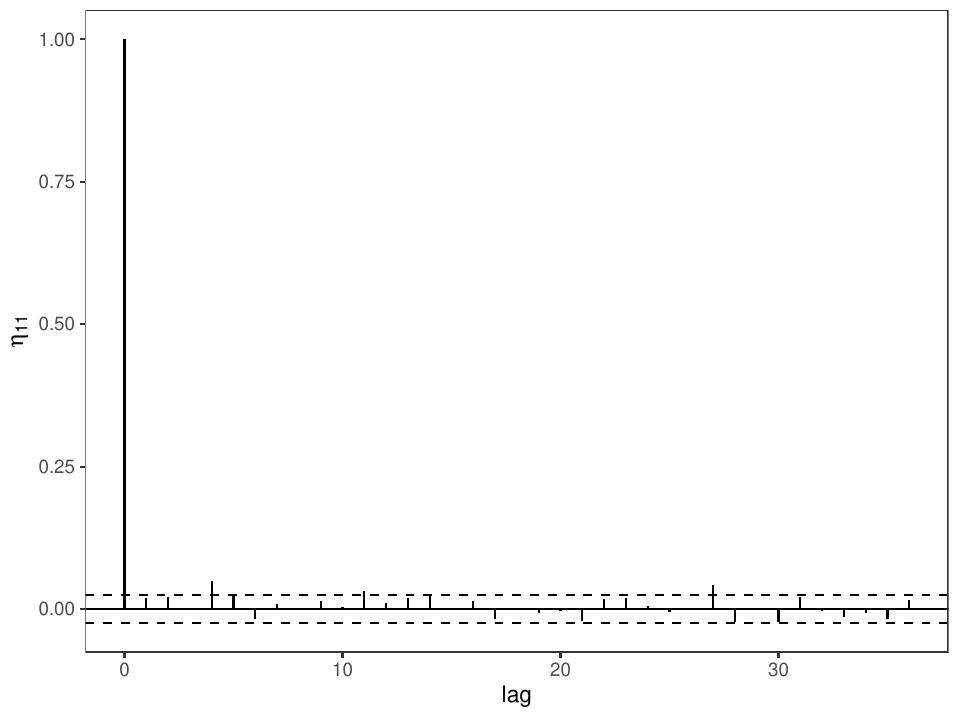}
\caption{The trace plots and autocorrelation plots for the posterior draws of $A_{1, 1}$ and $\eta_{1, 1}$.}
\label{fig:OAI-trace-acf}
\end{figure}

Figure~\ref{fig:OAI-trace-acf} presents the trace plots and autocorrelation plots for the posterior draws of $A_{1, 1}$ and $\eta_{1, 1}$. 
The trace plots indicate good convergence of the MCMC chains.
The draws of $\eta_{1, 1}$ appear nearly independent, 
while the autocorrelation plot of $A_{1, 1}$ decreases rapidly with increasing lag.
Similar plots for the remaining parameters are provided in the supplementary material.

\begin{figure}[ht]
\centering
\includegraphics[width = 0.95\textwidth]{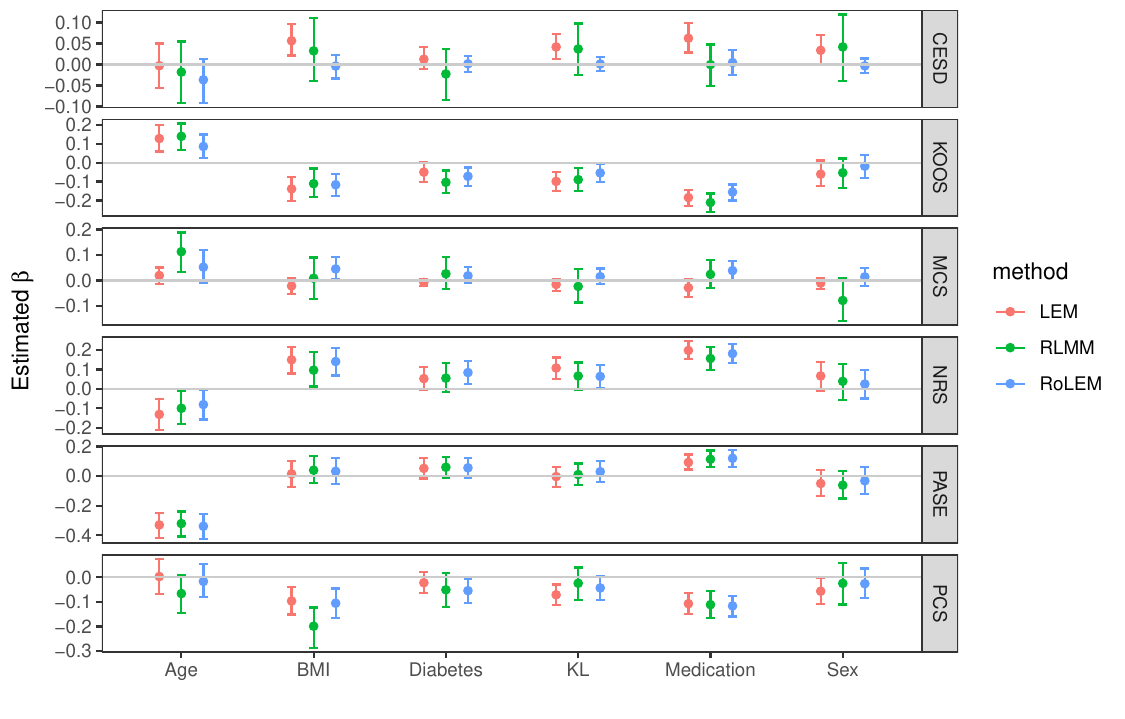}
\caption{The posterior means of $\beta$ and their associated 95\% HPD intervals obtained from LEM, RLMM, and RoLEM.}
\label{fig:OAI-beta}
\end{figure}

The posterior means of $\beta$ and their associated 95\% HPD intervals are shown in Figure~\ref{fig:OAI-beta}, together with the corresponding estimates obtained from LEM and RLMM. The numerical values are provided in the supplementary material.
The three methods yield similar estimates for some coefficients, but the lengths of the HPD intervals differ, reflecting varying levels of uncertainty.
For example, there is strong evidence that the presence of diabetes  is associated with higher NRS scores under RoLEM, whereas LEM and RLMM do not provide strong evidence for this association, as their 95\% HPD intervals include zero.

To compare the three models, LEM, RLMM, and RoLEM,  
Table~\ref{table:OAI-compare} reports their WAIC, BIC, and five-fold cross-validation (CV) scores. 
The CV performance is evaluated using both the mean log posterior density (MLPD) and the mean absolute error (MAE), where 
\begin{align*}
  \text{MLPD} = \frac{1}{n}\sum_{i=1}^n \log f_{\text{post}}(\tilde y_i), 
  \quad 
  \text{MAE} = \frac{1}{n}\sum_{i=1}^n \sum_{j=1}^J |\tilde y_{ij} - \hat y_{ij}|,
\end{align*}
where $\tilde y_i = (\tilde y_{i1}, \ldots, \tilde y_{iJ})$, $i = 1, \ldots, n$, denote the test data,
$f_{\text{post}}(\cdot)$ is the posterior predictive density and can be approximated using posterior draws, 
and $\hat y_{ij}$ denotes the posterior mean prediction. 
The results show that LEM performs worse than both RLMM and RoLEM, which is expected given the presence of outliers suggested by the trajectories in Figure~\ref{fig:OAI-y}, since LEM does not accommodate outliers. 
RLMM and RoLEM exhibit nearly identical cross-validation performance.
RLMM has a smaller WAIC, however, the difference in the WAIC is negligible compared with the magnitude of the WAIC values.
RoLEM attains a smaller BIC, and the difference can translate to a Bayes factor that indicating strong evidence in favor of RoLEM. 
Additionally, RoLEM consistently produces shorter HPD intervals than RLMM, as shown in Figure~\ref{fig:OAI-beta} and a plot in the supplementary material.
Thus, RoLEM offers a better performance over both LEM and RLMM.

\begin{table}[ht]
\centering
\caption{WAIC, BIC, and CV Scores for LEM, RLMM, and RoLEM}
\begin{tabular}{lcccc}
\hline 
Method & WAIC & BIC & CV-MLPD & CV-MAE \\
\hline 
LEM   & $14856.00$ & $14973.44$ & $-33.55$ & $0.74$ \\
RLMM  & $14162.97$ & $14393.41$ & $-32.00$ & $0.72$ \\
RoLEM & $14166.57$ & $14302.92$ & $-32.01$ & $0.72$ \\
\hline 
\end{tabular}
\label{table:OAI-compare}
\end{table}

We interpret the results based on the RoLEM estimates of $\beta$, 
which are shown in Figure~\ref{fig:OAI-beta} with numerical values provided in the supplementary material.
If a 95\% HPD interval excludes zero, there exists strong evidence of an association between the covariate and the response; otherwise, we conclude that no association is detected.
CESD measures depressive symptoms.
None of the covariates show strong evidence of association with CESD.
KOOS evaluates knee pain, with higher scores indicating less pain.
Higher age, lower BMI, absence of diabetes, lower KL grade, and no pain medication use are all associated with less knee pain.
MCS measures overall mental health status, with higher scores reflecting better mental health. 
Higher BMI is associated with better mental health.  
NRS measures non–activity-specific pain severity over the past 30 days, with higher scores indicating more pain.
Younger age, higher BMI, presence of diabetes, higher KL grade, and pain medication use are associated with more pain.
PASE assesses physical activity level, where higher scores reflect greater activity.
Younger age and pain medication use are associated with higher activity levels.
PCS measures physical health status, with higher values indicating better health.
Lower BMI, absence of diabetes, and no pain medication use are associated with better physical health.
Sex is not associated with any of the response variables.

Using the RoLEM, the posterior mean of $\rho$ is $0.51$ with a 95\% HPD interval of $(0.48, 0.53)$,
which indicates the existence of correlation among data collected from repeated visits for the same patients. 
The posterior mean of $\nu$, the degrees of freedom parameter, is $6.12$ with a 95\% HPD interval of $(4.78, 7.45)$, 
which provides evidence for the presence of outliers in the dataset.

Finally, we explore whether model fit could be improved by including pairwise interactions among the covariates.
For each candidate interaction, we added a single interaction term to the model and computed the corresponding WAIC and BIC values. The results are summarized in Table~\ref{table:OAI-interaction}.
Although adding certain interactions, for example, the BMI-medication interaction, yields models with smaller WAIC than the current model, none of the models with an interaction achieve a smaller BIC.
Recall that a smaller WAIC indicates better out-of-sample predictive performance, whereas comparing BIC corresponds to evaluating a Bayes factor.
Given that no model with interactions improves the BIC and the gains in WAIC are limited, we conclude that including interaction terms is not sufficiently beneficial and therefore do not pursue them further in this analysis.

\begin{table}[ht]
\centering
\caption{WAIC and BIC for Models with an Interaction}
\begin{tabular}{llccccc}
\hline 
  & & Age &  BMI &  Diabetes &  KL &  Medication \\
\hline 
WAIC
&BMI        & $14164.85$ &  \\
&Diabetes   & $14166.47$ & $14171.97$ &   \\
&KL         & $14171.53$ & $14163.59$ & $14172.79$ &   \\
&Medication & $14169.76$ & $14161.05$ & $14169.61$ & $14172.59$ & \\
&Sex        & $14171.77$ & $14171.97$ & $14166.57$ & $14162.28$ & $14170.04$ \\
\hline 

BIC
&BMI        & $14304.78$ &   \\
&Diabetes   & $14310.60$ & $14310.82$ &   \\
&KL         & $14314.39$ & $14305.87$ & $14311.12$ &   \\
&Medication & $14312.12$ & $14303.06$ & $14314.86$ & $14312.32$ &   \\
&Sex        & $14314.32$ & $14314.36$ & $14302.92$ & $14304.60$ & $14308.64$ \\
\hline 
\end{tabular}
\label{table:OAI-interaction}
\end{table}

\section{Conclusion}
\label{section:conclusion}

This paper proposes RoLEM, which extends existing methods in three key aspects:
(1) modeling the correlation structure among repeated measurements from the same subject;
(2) modeling random errors using a scale mixture of matrix-variate normal distributions to accommodate potential outliers; and
(3) introducing novel prior and proposal distributions for the envelope model.
A central challenge in the Bayesian inference for RoLEM is the parameterization of $\Sigma_{\varepsilon}$ and the modeling of $\Gamma$ and $\Gamma_0$. We address this by working directly with the projection matrix corresponding to the subspace spanned by $\Gamma$ and imposing a prior distribution on the Grassmann manifold.
Both the simulation studies and real data analysis demonstrate the superior performance of RoLEM.


The model  relies on certain intrinsic assumptions. In the most general setting, the variance matrix of $\mathrm{vec}(Y_i)$ is an $rJ_i \times rJ_i$ matrix. However, RoLEM assumes that it admits the structured form $R_i \otimes \Sigma_{\varepsilon}$, which implies that the variance of each response is constant across time points and that the temporal correlation structure stays the same for all responses. The model also assumes that the kurtosis is identical across all outcomes.
When this homogeneous structure does not hold, more flexible and complex variance structures may be required. 
If the variances differ across time, we may introduce a time-specific error covariance $\Sigma_{\varepsilon,j}$ for the $j$th time point. 
In this case, we can impose an envelope structure of the form
$\Sigma_{\varepsilon, j} = \Gamma\Omega_j\Gamma^T + \Gamma_0\Omega_{0, j}\Gamma_0^T$,
which maintains a common envelope basis across time while allowing $\Omega_j$ and $\Omega_{0,j}$ to vary with time $j$. Under this structure, we may still assume $\beta = \Gamma \eta$.
We will investigate this issue further in future research.

Missing data are also common in longitudinal studies. For simplicity, we remove subjects with incomplete observations. In practice, however, various imputation methods can be applied prior to fitting RoLEM. 

\color{black}






\bibliographystyle{plain}
\bibliography{references}

@article{schwarz1978estimating,
  title={Estimating the dimension of a model},
  author={Schwarz, Gideon},
  journal={The Annals of Statistics},
  pages={461--464},
  year={1978},
  publisher={JSTOR}
}

@article{watanabe2010asymptotic,
  title={Asymptotic equivalence of {B}ayes cross validation and widely applicable information criterion in singular learning theory},
  author={Watanabe, Sumio},
  journal={Journal of Machine Learning Research},
  volume={11},
  number={12},
  pages={3571--3594},
  year={2010}
}

@article{kass1995bayes,
  title={Bayes factors},
  author={Kass, Robert E and Raftery, Adrian E},
  journal={Journal of the American Statistical Association},
  volume={90},
  number={430},
  pages={773--795},
  year={1995},
  publisher={Taylor \& Francis}
}

@book{gelman2013bayesian,
  title={Bayesian Data Analysis},
  author={Gelman, Andrew and Carlin, John B and Stern, Hal S and Dunson, David B
  and Vehtari, Aki and Rubin, Donald B},
  year={2013},
  edition={3rd},
  publisher={Chapman and Hall/CRC}
}

@article{khare2017bayesian,
  title={A {B}ayesian approach for envelope models},
  author={Khare, Kshitij and Pal, Subhadip and Su, Zhihua},
  journal={The Annals of Statistics},
  volume={45},
  number={1},
  pages={196--222},
  year={2017}
}

@article{chakraborty2023comprehensive,
  title={A comprehensive {B}ayesian framework for envelope models},
  author={Chakraborty, Saptarshi and Su, Zhihua},
  journal={Journal of the American Statistical Association},
  volume={119},
  number={547},
  pages={2129--2139},
  year={2024}
}

@book{GuptaNagar2000,
  author    = {Gupta, A. K. and Nagar, D. K.},
  title     = {Matrix Variate Distributions},
  year      = {2000},
  publisher = {Chapman \& Hall/CRC},
  address   = {Boca Raton},
}

@article{wang2012bayesian,
  title={Bayesian analysis of multivariate $t$-linear mixed models
         using a combination of {IBF} and {G}ibbs samplers},
  author={Wang, Wan-Lun and Fan, Tsai-Hung},
  journal={Journal of Multivariate Analysis},
  volume={105},
  number={1},
  pages={300--310},
  year={2012},
  publisher={Elsevier}
}

@book{chikuse2003statistics,
  title={Statistics on Special Manifolds},
  author={Chikuse, Yasuko},
  year={2003},
  publisher={Springer}
}

@article{Eitner2021,
  author  = {Eitner, Annett and Culvenor, Adam G. and Wirth, Wolfgang and Schaible, Hans-Georg and Eckstein, Felix},
  title   = {Impact of Diabetes Mellitus on Knee Osteoarthritis Pain and Physical and Mental Status: Data From the Osteoarthritis Initiative},
  journal = {Arthritis Care \& Research},
  year    = {2021},
  volume  = {73},
  number  = {4},
  pages   = {540--548},
  doi     = {10.1002/acr.24173}
}

@book{azzalini2014skew,
  title={The Skew-Normal and Related Families},
  author={Azzalini, Adelchi and Capitanio, Antonella},
  year={2014},
  publisher={Cambridge University Press}
}

@article{cook2010envelope,
  title={Envelope models for parsimonious and efficient multivariate linear regression},
  author={Cook, R Dennis and Li, Bing and Chiaromonte, Francesca},
  journal={Statistica Sinica},
  pages={927--960},
  year={2010},
  volume={20},
  number={3},
  publisher={JSTOR}
}

@article{cook2015foundations,
  title={Foundations for envelope models and methods},
  author={Cook, R Dennis and Zhang, Xin},
  journal={Journal of the American Statistical Association},
  volume={110},
  number={510},
  pages={599--611},
  year={2015},
  publisher={Taylor \& Francis}
}

@article{ding2018matrix,
  title={Matrix variate regressions and envelope models},
  author={Ding, Shanshan and Cook, R Dennis},
  journal={Journal of the Royal Statistical Society Series B: Statistical Methodology},
  volume={80},
  number={2},
  pages={387--408},
  year={2018},
  publisher={Oxford University Press}
}

@article{ding2021envelope,
  title={Envelope quantile regression},
  author={Ding, Shanshan and Su, Zhihua and Zhu, Guangyu and Wang, Lan},
  journal={Statistica Sinica},
  volume={31},
  number={1},
  pages={79--105},
  year={2021},
  publisher={JSTOR}
}

@article{fonseca2008objective,
  title={Objective Bayesian analysis for the Student-t regression model},
  author={Fonseca, Tha{\'\i}s C O and Ferreira, Marco A R and Migon, Helio S},
  journal={Biometrika},
  volume={95},
  number={2},
  pages={325--333},
  year={2008},
  publisher={Oxford University Press}
}

@article{ding2014bayesian,
  title={Bayesian robust inference of sample selection using selection-$t$ models},
  author={Ding, Peng},
  journal={Journal of Multivariate Analysis},
  volume={124},
  pages={451--464},
  year={2014},
  publisher={Elsevier}
}

@article{fernandez1999multivariate,
  title={Multivariate Student-$t$ regression models: Pitfalls and inference},
  author={Fern{\'a}ndez, Carmen and Steel, Mark F J},
  journal={Biometrika},
  volume={86},
  number={1},
  pages={153--167},
  year={1999},
  publisher={Oxford University Press}
}

@article{lee2020review,
  title={A review of envelope models},
  author={Lee, Minji and Su, Zhihua},
  journal={International Statistical Review},
  volume={88},
  number={3},
  pages={658--676},
  year={2020},
  publisher={Wiley Online Library}
}

@book{cook2018introduction,
  title={An Introduction to Envelopes: Dimension Reduction for Efficient Estimation in Multivariate Statistics},
  author={Cook, R Dennis},
  year={2018},
  publisher={John Wiley \& Sons}
}

@article{shi2020mixed,
  title={Mixed effects envelope models},
  author={Shi, Yuyang and Ma, Linquan and Liu, Lan},
  journal={Stat},
  volume={9},
  number={1},
  pages={e313},
  year={2020},
  publisher={Wiley Online Library}
}

\end{document}